\documentclass[pdflatex,sn-mathphys]{sn-jnl}

\jyear{2023}%

\theoremstyle{thmstyleone}%
\theoremstyle{thmstyletwo}%

\theoremstyle{thmstylethree}%

\usepackage{subfigure}

\usepackage{color}
\usepackage{ulem}
 
\begin{document}

\title[Quantum and higher curvature corrections to the anti-de Sitter black hole]{Quantum and higher curvature corrections to the anti-de Sitter black hole}


\author[1,2]{\fnm{Kristian Hauser} \sur{Villegas}}\email{kavillegas1@up.edu.ph}
\equalcont{These authors contributed equally to this work.}

\author[3]{\fnm{Reginald Christian} \sur{Bernardo}}\email{rbernardo@gate.sinica.edu.tw}
\equalcont{These authors contributed equally to this work.}

\affil[1]{\orgdiv{National Institute of Physics}, \orgname{University of the Philippines Diliman}, \orgaddress{\city{Quezon City}, \postcode{1101}, \country{Philippines}}}

\affil[2]{\orgdiv{Division of Physics and Applied Physics}, \orgname{Nanyang Technological University}, \orgaddress{\city{Singapore}, \postcode{637371}, \country{Singapore}}}

\affil[3]{\orgdiv{Institute of Physics}, \orgname{Academia Sinica}, \orgaddress{\city{Taipei}, \postcode{11529}, \country{Taiwan}}}




\abstract{Black holes exert quantum pressure coming from the nonlocal gravity correction. We investigate this nonlocal correction for black holes in anti-de Sitter (AdS) spacetime and its dual boundary field theory. We show that the second order curvature and the nonlocal actions do not backreact on the AdS black hole metric. Thus, the interpretation of quantum pressure holds in the bulk for AdS black hole, generalizing the previous result for the asymptotically flat black hole. We then show that the leading geometric correction comes from the third order in curvature and explicitly calculate the corrections to the metric and to the horizon. For applications to AdS/CFT, we conjectured a nonlocal Gibbons-Hawking-York boundary term along with the necessary counter terms to cancel the ultraviolet divergence of the bulk action. We then calculate the thermodynamic quantities in the bulk and discuss their properties.}




\maketitle


\section{Introduction}
Using classical gravity to understand a quantum system via holographic principle is one of the most profound discovery of contemporary theoretical physics \cite{Maldacena1998}. It has applications in various fields from quantum chromodynamics to condensed matter physics \cite{Kovtun2005, Heinz2012, Hartnoll2008, Zaanen2015}. In addition, there are also attempts to understand the bulk black hole by studying the dual boundary theory \cite{Nosaka2020, Lensky2021}. Extensions of the bulk Einstein gravity have also been considered, which was shown to lead to different universality classes of the boundary conformal field theories \cite{Myers2010, Parvizi2019}. Crucial to this field is the anti de Sitter (AdS) spacetime where the holographic duality is best understood.

It is known since the 1970s \cite{Bekenstein1973,Hawking1975} that black holes obey the laws of thermodynamics \cite{Wald2001}, which has been taken as a hint that gravity is an emergent property from an underlying microscopic quantum gravity. In the usual formulation however, the pressure and volume conjugate variables are missing. There are proposals to include such variables \cite{Kastor:2009wy, Kastor:2010gq, Kastor:2011qp, Dolan2011a,Dolan2011b}. Here, the pressure comes from the nonzero cosmological constant and the conjugate volume is related to the volume of the bulk black hole in zero angular momentum limit. In the anti de Sitter/conformal field theory (AdS/CFT) correspondence, the extra dimension that forms the AdS bulk comes from the geometrization of the renormalization group structure of the dual field theory. One would then expect that the volume conjugate to the pressure of the dual field theory should be the volume only of the boundary surface. It is not clear, therefore, if the pressure coming from the cosmological constant is the correct pressure of the dual field theory. In fact, in the usual zero density AdS/CFT applications, there is no pressure in the thermodynamics relations \cite{Zaanen2015}.

Recently, it was shown that the nonlocal quantum corrections to the Einstein gravity leads to the quantum pressure in black holes \cite{Calmet:2021lny}. Such an interpretation depends on the fact that in asymptotically flat black holes the metric does not receive corrections from the nonlocal and quadratic curvature terms. Given the importance of the holographic duality applications in various fields in physics, this motivates us to ask the following questions: Does the nonlocal action also gives rise to quantum pressure for AdS spacetime? How are the thermodynamic quantities such as entropy, temperature, and pressure modified? What is the corresponding nonlocal boundary action?

We investigate these questions and show that there are no corrections to the AdS metric coming from the nonlocal and second order curvature terms, thus showing that the nonlocal action gives rise to the quantum pressure in the bulk. We then derive the appropriate boundary terms and calculate the thermodynamic quantities.

We bear in mind that in \cite{Pourhassan:2022auo} the thermodynamic quantities for the electrically charged AdS black hole were calculated. This hinges on a nonvanishing nonlocal metric correction that is second order in the gravitational field strength. However, in contrast, we do not obtain any metric correction to the AdS black hole up to second order in the curvature and nonlocal physics, as we were expressing our results with respect to the curvature. In this way, we were able to generalize the notion of the quantum pressure to an AdS black hole. We additionally show that third order curvature sources a metric correction, influencing the event horizon and by extension the thermodynamics anchored on it.

This paper is organized as follows. We start by briefly reviewing how the quantum pressure is obtained from the nonlocal action (Section \ref{quantumpressure}). In Section \ref{ads} we consider the AdS spacetime. Following the logic that lead to the interpretation of quantum pressure as a nonlocal effect, we show that the quadratic curvature and nonlocal actions do not change the metric and hence the black hole horizon. In this case, since the AdS metric is unaffected, the quantum pressure can only be associated with the nonlocal action, thereby generalizing its notion beyond asymptotic flatness. In addition, we calculate the third-order curvature correction to the metric. This is shown to reduce to the known Schwarzschild solution \cite{Calmet:2021lny} in the asymptotically flat limit. In Section \ref{boundary} we derive the boundary term for the action. We then calculate the thermodynamic quantities in Section \ref{thermo}. Lastly, we give our conclusions.

We work with the mostly plus metric signature, $(-, +, +, +)$, and geometrized units, $c = G_\text{N} = 1$, where $c$ and $G_\text{N}$ are the speed of light in vacuum and Newton's gravitational constant. Mathematica and python notebooks which can be used to reproduce the results of this paper can downloaded from \href{https://github.com/reggiebernardo/notebooks/tree/main/supp_ntbks_arxiv.2208.07663}{GitHub} \cite{reggie_bernardo_4810864}.

\section{Quantum pressure}
\label{quantumpressure}
In this section we briefly review how the black hole pressure emerges from the quantum correction following \cite{Calmet:2021lny}. The local effective action up to quadratic curvature order is given by
\begin{align}
\label{localaction0}
S= 
\int d^4x \sqrt{-g} \bigg( \kappa R 
& +c_1(\mu)R^2 +c_2(\mu)R_{\mu\nu}R^{\mu\nu} + c_3(\mu)R_{\mu\nu\rho\sigma}R^{\mu\nu\rho\sigma} \bigg) \,,
\end{align}
where $\kappa = M_\text{Pl}^2/2 = 1/ \left(16 \pi G_\text{N} \right)$ with $G_\text{N}$ the gravitational constant, $M_\text{Pl} \approxeq 2.4 \times 10^{18}$ GeV is the reduced Planck mass, and $\mu$ is the renormalization scale. The dependence of the coefficients $c_i$'s on $\mu$ simply means that they are sensitive to the cut-off scale if we try to calculate them from first principles \cite{Barvinsky1983}. These coefficients are dimensionless numbers to be fixed by the experiment \cite{Donoghue1994}. The first term in \eqref{localaction0} is the Einstein-Hilbert term, which is trailed by the quadratic curvature operators. The quantum pressure comes from the nonlocal quantum gravity contribution
\begin{align}
\label{nonlocalaction}
S_\text{nl}=-\gamma\int d^4x \sqrt{-g} R_{\mu\nu\alpha\beta}\ln \bigg(\frac{\Box}{\mu^2}\bigg)R^{\mu\nu\alpha\beta} \,,
\end{align}
where $\gamma$ characterizes the strength of the nonlocal interaction. The correction to the horizon comes from the third order in curvature \cite{Calmet2017}
\begin{align}
\label{eq:O3term}
S_3=\lambda_3 \int d^4x \sqrt{-g} R^{\mu\nu}_{\;\;\alpha\sigma}R^{\alpha\sigma}_{\;\;\delta\gamma}R^{\delta\gamma}_{\;\;\mu\nu} \,.
\end{align}

The Wald entropy formula
\begin{align}
\label{waldentropy}
\mathcal{S}_{\text{Wald}}=&-2\pi\int d\Sigma \epsilon_{\mu\nu}\epsilon_{\rho\sigma}\frac{\delta\mathcal{L}}{\delta R_{\mu\nu\rho\sigma}}\bigg\vert_{r=r_H}
\end{align}
can be used to obtain \cite{Calmet:2021lny}
\begin{align}
\label{waldentropy2}
\mathcal{S} =\frac{A}{4G_\text{N}} &+64\pi^2c_3(\mu)+64\pi^2\gamma[\log (4G_\text{N}^2 \mathcal{M}^2\mu^2)-2+2\gamma_\text{E}] \,,
\end{align}
which is valid up to quadratic order in the curvature, where $\mathcal{M}$ is the black hole mass and $\gamma_\text{E}$ is Euler's constant. The first term is the familiar Bekenstein-Hawking entropy where $A$ is the black hole area, whereas the correction terms give rise to quantum pressure. This gives the thermodynamic relation
\begin{align}
\label{thermo}
TdS-PdV=\bigg(1+\frac{16\pi\gamma}{G_\text{N} \mathcal{M}^2}\bigg)d\mathcal{M}
\end{align}
and the quantum pressure
\begin{align}
\label{pressure1}
P=-\frac{\gamma}{2G_\text{N}^4 \mathcal{M}^2} \,.
\end{align}

As emphasized in \cite{Calmet:2021lny}, the interpretation of \eqref{pressure1} as pressure comes from the fact that the metric, and hence the horizon, does not receive correction in second-order curvature expansion from the quadratic EFT and nonlocal terms.

It is important to point out that the results in \cite{Calmet:2021lny} were obtained considering asymptotic flatness. In this work, we depart from this assumption and look for the leading order corrections to the nonlocal physics due to the presence of an AdS boundary, with the AdS length scale acting as a perturbation parameter. We therefore also take into account the additional nonlocal term where the Ricci tensor appears in \eqref{nonlocalaction} instead of the Riemann tensor \cite{Calmet:2021lny, Calmet:2018elv}. We will find that such contributions are subdominant, as expected by a factor inversely proportional to the AdS length scale, compared to the nonlocal effects anchored on \eqref{nonlocalaction}.

\section{EFT and nonlocal corrections to the AdS metric}
\label{ads}
Although the holographic principle is now believed to be much more general, the well-understood duality is the AdS/CFT where the bulk has anti-de Sitter spacetime \cite{Gubser1998, Witten1998}. In this section we will calculate the corrections to the Schwarzschild-AdS metric due to the higher order curvature and nonlocal actions. We will show that there are no corrections from second order curvature and from leading order nonlocal actions. We explicitly calculate the third order curvature perturbation to the metric, which is the lowest nontrivial correction.

\subsection{Quadratic curvature correction}

The quadratic curvature effective field theory (EFT) action \eqref{localaction0} becomes
\begin{align}
\label{localaction}
S= 
& \int d^4x \sqrt{-g} \bigg(\kappa \left( R - 2\Lambda \right)
+c_1(\mu)R^2  +c_2(\mu)R_{\mu\nu}R^{\mu\nu}+c_3(\mu)R_{\mu\nu\rho\sigma}R^{\mu\nu\rho\sigma} \bigg) \,,
\end{align}
where $\Lambda = {\cal O}\left(L^{-2}\right)$ is a cosmological constant and $L$ is the AdS length scale. Since we are mostly interested in the influence of the AdS boundary on the bulk and thermodynamic properties, we refer to the AdS length scale as a reference, and calculate the leading order EFT and nonlocal corrections with respect to this ruler. We do so first by showing that the quadratic curvature EFT do not change the AdS metric. 

The equation of motion can be obtained by varying the action with respect to the metric. Starting with the Einstein-Hilbert term together with the cosmological constant, we obtain the well known
\begin{equation}
\label{eq:eomeh}
\kappa \left( G_{\beta\gamma} + \Lambda g_{\beta \gamma} \right) \,.
\end{equation}
Now, we deal with the quadratic curvature EFT terms. The terms proportional to $c_1\left(\mu\right)$ lead to
\begin{equation}
\label{eq:eomquad1}
    c_1\left(\mu\right)\left( 2 G_{\beta \gamma} R + \dfrac{1}{2} g_{\beta\gamma} R^2 + 2 g_{\beta\gamma} \Box R - 2 \nabla_\gamma \nabla_\beta R \right) \,,
\end{equation}
the terms proportional to $c_2\left(\mu\right)$ give
\begin{equation}
\label{eq:eomquad2}
\begin{split}
    c_2\left(\mu\right) \bigg( & 2 G_{\beta}^{\ \alpha} G_{\gamma \alpha} - \dfrac{g_{\beta\gamma}}{2} G_{\alpha\delta}G^{\alpha\delta} + G_{\beta \gamma} R 
    - \dfrac{g_{\beta\gamma}}{2} G_\alpha^{\ \alpha} R \\
    & + G_\beta^{\ \alpha} g_{\gamma\alpha} R + \Box G_{\beta \gamma} 
    + g_{\beta\gamma} \Box R - 2 \nabla_\alpha \nabla_{(\beta} G_{\gamma)}^{\ \alpha} - \nabla_\gamma \nabla_\beta R \bigg) \,,
\end{split}
\end{equation}
and lastly the terms attached to $c_3\left(\mu\right)$ give\footnote{Although, we note that the easier path to solving the field equations is by using the Gauss-Bonnet identity to eliminate the Kretschmann scalar term, $R_{\mu\nu\rho\sigma}R^{\mu\nu\rho\sigma}$, in the action before variation.}
\begin{equation}
\label{eq:eomquad3}
\begin{split}
    c_3\left(\mu\right) \bigg( & - \dfrac{g_{\beta\gamma}}{2} R_{\alpha\delta\mu\nu}R^{\alpha\delta\mu\nu} + 2 R_\beta^{\ \delta\mu\nu} R_{\gamma\delta\mu\nu} + 4 \nabla_\delta \nabla_\alpha R_{\beta \ \gamma \ }^{\ \alpha \ \delta} \bigg) \,.
\end{split}
\end{equation}

The above terms in the field equations can be each shown to vanish when evaluated on the AdS \cite{Tsai:2011gv, Charmousis:2019fre} black hole with
\begin{equation}
    \Lambda = -\dfrac{3}{L^2} \,.
\end{equation}
This can be verified with the following useful identities derived from the AdS solution:
\begin{eqnarray}
    R_{\alpha \beta} &=& \Lambda g_{\alpha \beta} \,, \nonumber \\
    R &=& 4 \Lambda \,, \nonumber \\
    G_{\alpha \beta} &=& -\Lambda g_{\alpha \beta} \,, \nonumber \\
    R_{\mu\nu\sigma\tau}R^{\mu\nu\sigma\tau} &=& \dfrac{8}{3} \Lambda^2  + \dfrac{48{\cal M}^2}{r^6}  \,, \nonumber \\
    \label{maxsymidentities} R_\alpha^{\ \mu\nu\sigma} R_{\beta \mu\nu\sigma} &=& \dfrac{2}{3} \Lambda^2 g_{\alpha \beta} + {\cal C}_{\alpha \beta}\left({\cal M}, r\right) \,,
\end{eqnarray}
where ${\cal M}$ and $r$ are the Schwarzschild mass and radial coordinate, respectively, and $C_{\alpha\beta}\left({\cal M}, r\right)$ are the components of $R_\alpha^{\ \mu\nu\sigma} R_{\beta \mu\nu\sigma}$ in the nonrotating, asymptotically flat limit\footnote{Clearly in the asymptotically flat limit, $L \rightarrow \infty$ or equivalently $\Lambda \rightarrow 0$, the above relations reflect the well known Ricci flat metric conditions satisfied by the Kerr solution.}. The AdS black hole with the AdS scale $L$ is therefore an exact solution of the quadratic curvature EFT. We emphasize that the proof of this relies solely on the first three \textit{covariant} equations of \eqref{maxsymidentities} that are sufficient to make \eqref{eq:eomquad1} and \eqref{eq:eomquad2} vanish, i.e., the $c_3\left(\mu\right)$ terms can be absorbed into \eqref{eq:eomquad1} and \eqref{eq:eomquad2} by using the Gauss-Bonnet identity. We were merely showing \eqref{eq:eomquad3} and the fourth and fifth lines of \eqref{maxsymidentities} for completeness, but do not rely on them for the derivation of our results.

In the nonrotating limit, the line element of the spacetime can be written as
\begin{equation}
\label{eq:ss_ansatz}
ds^2=\frac{r^2}{L^2}\left(-f(r)dt^2 + L^2 d\Omega^2 \right)+\frac{L^2}{r^2f(r)}dr^2 \,,
\end{equation}
where $d\Omega^2 = d\theta^2 + \sin^2 \theta d\varphi^2$ and the metric function is given by
\begin{equation}
\label{ef1}
    f(r) = 1 + \dfrac{L^2}{r^2} - \dfrac{M^3}{r^3} \,.
\end{equation}
Above, $M$ is an integration constant with the dimensions of length, not to be confused with the black hole mass $\mathcal{M}$.

The black hole event horizon is located at
\begin{equation}
\label{horizonfull}
\begin{split}
    r_H = \ & \dfrac{\left( \sqrt{12 L^6+81 M^6}+9 M^3 \right)^{1/3}}{2^{1/3} 3^{2/3}} - \left( \frac{2}{3} \right)^{1/3} \dfrac{ L^2}{\left(\sqrt{12 L^6+81 M^6}+9 M^3\right)^{1/3}} \,.
\end{split}
\end{equation}
The integration constant $M$ gives the length scale of the black hole, which, as will be shown explicitly below, is proportional to the horizon radius. The two relevant length scales, the AdS boundary scale $L$ and the black hole length scale $M$, give rise to two interesting limits, a large black hole ($L/M \ll 1$), when the black hole is larger than the AdS boundary, and a small black hole ($L/M \gg 1$), when the AdS boundary is too far away from the black hole.

In the large black hole limit, we have $L/M \ll 1$ and find
\begin{equation}
\label{horizonlarge}
r_H = M - L^2/(3 M) + O \left( \left( L/M \right)^3 \right).
\end{equation}
This shows that in the extreme limit the black hole length scale $M$ itself becomes the event horizon radius $r_H$. On the other end, for a small black hole, or alternatively the Schwarzschild limit, $L/M \gg 1$, we get 
\begin{equation}
\label{horizonschwarszchild}
r_H = M^3/L^2 + O \left( \left( M/L \right)^8 \right) \,.
\end{equation}
This limit permits the identification of the Schwarzschild mass $\mathcal{M}$ in terms of the black hole and AdS scales to be 
\begin{equation}
\label{eq:Mschw}
\mathcal{M} = M^3/\left(2L^2\right)   
\end{equation}
such that the familiar relation appears
\begin{equation}
r_H \sim 2\mathcal{M} \ \ \ \ \text{as} \ \ \ \ M/L \rightarrow 0 \,.
\end{equation}

The large and small AdS black hole limits have quite interesting physical properties \cite{Cardoso:2001bb, Cardoso:2003cj, Wang:2021uix, Wang:2021upj, Fortuna:2022fdd}. The large black hole present an intriguing electromagnetic quasinormal spectrum featuring bifurcation profile and overdamped modes. On the other hand, the small AdS black hole predicts the first signatures, ${\cal O}\left(1/L^2\right)$, we may expect, if any, from the existence of a spacetime boundary in an astrophysical setting. We will also later show that these limiting cases correspond to the high and low temperature limits of the black hole.

\begin{figure}[h!]
\center
	\subfigure[  ]{
		\includegraphics[width = 0.7 \textwidth]{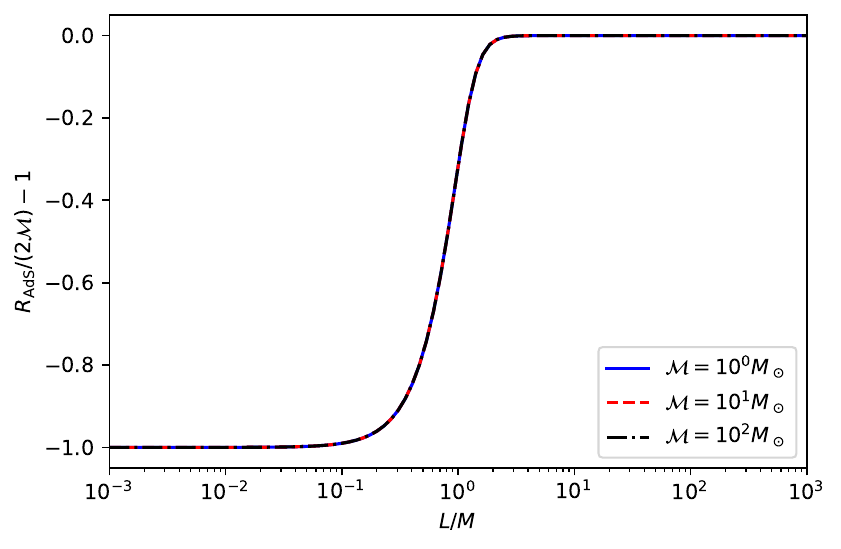}
		}
	\subfigure[  ]{
		\includegraphics[width = 0.7 \textwidth]{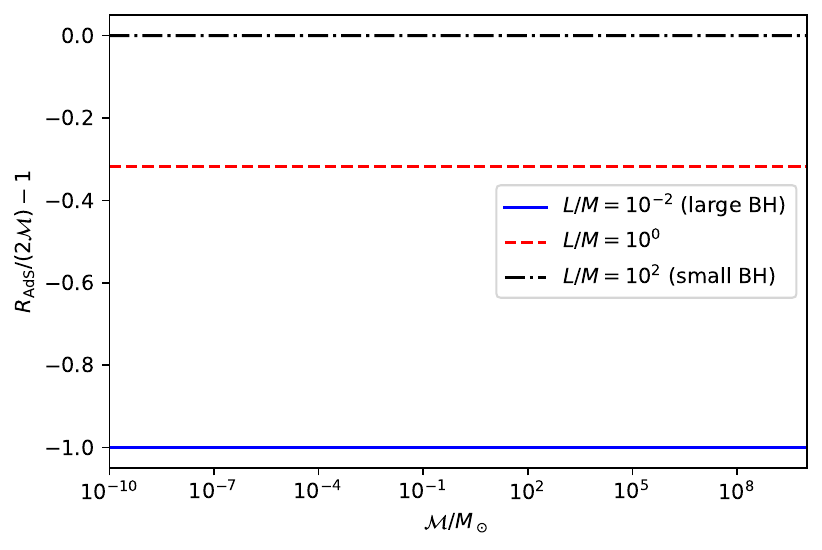}
		}
\caption{AdS black hole event horizon radius, $R_{\rm AdS} = r_H$, dependencies on (a) $L/M$ and (b) $\cal M$.}
\label{fig:AdShorizon}
\end{figure}

Before we leave this section, we visualize below the AdS black hole's event horizon radius (Figure \ref{fig:AdShorizon}).

We find that for a fixed mass $\cal M$, varying $L/M$ confirms the small and large black hole limits we have discussed. This is transparent in Figure \ref{fig:AdShorizon}(a), where $R_{AdS}/(2{\cal M}) - 1$ vanishes in the small black hole limit $L/M \gg 1$ as the event horizon radius reduces to that of the Schwarzschild, $r_H = 2{\cal M}$. On the other hand, in the large black hole limit $L/M \ll 1$, we see that $R_{\rm AdS} \ll 2 \cal M$, which we can understand through \eqref{horizonlarge} and \eqref{eq:Mschw}, i.e., $R_{\rm AdS}/(2{\cal M}) \sim M/(2{\cal M}) = (L/M)^2 \ll 1$ as $L/M \ll 1$. The intermediate regime between the large and small black hole cases is also shown, revealing a smooth transition between the extreme cases.

We emphasize that the mass $\cal M$ is only meaningful in the small black hole or Schwarzschild limit. Nonetheless, we use it in the plots as it is a useful reference, particularly for possible astrophysical applications. We find on the other hand that for fixed $L/M$ and varying $\cal M$ that the relative position of the event horizon is unaltered (Figure \ref{fig:AdShorizon}(b)) regardless of a small, intermediate, or large black hole. This could have also be predicted based on Figure \ref{fig:AdShorizon}(a) where the different curves corresponding to various masses $\cal M$ simply overlapped and appeared visually indistinct. We move on the next section to see that this geometry holds even with nonlocal effects.

\subsection{Nonlocal metric correction}
\label{subsec:nonlocalmetric}

In \cite{Calmet:2021lny}, it was shown that the nonlocal action does not change the Schwarzschild metric. We generalize this result in this section by showing that the nonlocal action still does not change the form of the AdS black hole. That is, the AdS black hole (\eqref{eq:ss_ansatz} and \eqref{ef1}) is also a solution to the leading order nonlocal sector of the theory.

To see this, we show that the nonlocal action \eqref{nonlocalaction} vanishes on the AdS solution and so does not change the extremum of the action upon which this classical AdS phase space trajectory holds. As in \cite{Calmet:2021lny}, we resort to the nonlocal Gauss-Bonnet identity
\begin{equation}
\label{gaussbonnet}
\begin{split}
    R_{\mu\nu\rho\sigma} \ln \left( \dfrac{\Box}{\mu^2} \right) R^{\mu\nu\rho\sigma} = \ & 4 R_{\mu\nu} \ln \left( \dfrac{\Box}{\mu^2} \right) R^{\mu\nu} - R \ln \left( \dfrac{\Box}{\mu^2} \right) R + O \left( R^3 \right) \,,
\end{split}
\end{equation}
where ${\cal O}\left(R^3\right)$ encloses at least third order curvature terms. In terms of the AdS boundary scale $L$, these third order curvature terms correspond to ${\cal O}\left(L^{-6}\right)$. Then, the nonlocal action \eqref{nonlocalaction} becomes
 \begin{align}
\label{nonlocalaction2}
S_\text{nl}=-\gamma\int d^4x \sqrt{-g} \bigg[ & 4R_{\mu\nu}\ln \bigg(\frac{\Box}{\mu^2}\bigg)R^{\mu\nu} -R\ln \bigg(\frac{\Box}{\mu^2}\bigg)R\bigg]  + O \left( R^3 \right)\,.
 \end{align}
We then use the representation \cite{Barvinsky1990}
\begin{align}
\label{integralrep}
\ln \frac{\Box}{\mu^2}=\int_0^\infty ds\bigg(\frac{1}{\mu^2+s}-\frac{1}{\Box+s}\bigg) \,,
\end{align}
along with the identities in \eqref{maxsymidentities}, to deal with the nonlocal operations. The first integrand becomes
\begin{align}
4R_{\mu\nu}\ln \bigg(\frac{\Box}{\mu^2}\bigg)R^{\mu\nu}=\int_{m_\text{IR}^2}^\infty ds\bigg(\frac{16\Lambda^2}{\mu^2+s}-\frac{16\Lambda^2}{s}\bigg) \,,
\end{align}
where we used $\Box g_{\mu\nu}=0$ coming from metric compatibility with the curvature. Note that we added a threshold $m^2_\text{IR}$ for the lower bound of the integral to control the infrared divergence. Performing the integration we have
\begin{align}
4R_{\mu\nu}\ln \bigg(\frac{\Box}{\mu^2}\bigg)R^{\mu\nu}=16\Lambda^2\ln\frac{m^2_\text{IR}}{\mu^2+m^2_\text{IR}} \,.
\end{align}
Similar manipulations yield
\begin{align}
\label{RRterm}
R\ln \bigg(\frac{\Box}{\mu^2}\bigg)R=16\Lambda^2\ln\frac{m^2_\text{IR}}{\mu^2+m^2_\text{IR}} \,.
\end{align}
Substituting these into \eqref{nonlocalaction2} we see that there is a cancellation and this gives
\begin{align}
S_\text{nl} = 0
\end{align}
when the nonlocal action is evaluated on the AdS metric (\eqref{eq:ss_ansatz} and \eqref{ef1}).

This confirms that the nonlocal action \eqref{nonlocalaction} does not add value to the action on the AdS metric. In fact, the Kerr-AdS metric continues to be solution to the field equations in the presence of the nonlocal terms. This generalizes the previous result (in the Schwarzschild limit \cite{Calmet:2021lny}): `that the quadratic curvature and nonlocal terms are solved by the Kerr-AdS black hole'.

To further confirm that the nonlocal action does not give rise to corrections in the solution, we perform the variation of \eqref{nonlocalaction} with respect to the metric. This gives two main contributions
\begin{equation}
\label{s1ands2}
    \delta S_\text{nl} = \delta S_1 + \delta S_2 \,,
\end{equation}
where
\begin{equation}
\label{s1}
    \dfrac{ \delta S_1 }{\gamma} = \int d^4 x \dfrac{\sqrt{-g}}{2} g_{\sigma \tau} \delta g^{\sigma \tau} R_{\mu\nu\alpha\beta} \ln \left( \dfrac{\Box}{\mu^2} \right) R^{\mu\nu\alpha\beta} \,
\end{equation}
and
\begin{equation}
\label{s2}
    -\dfrac{ \delta S_2 }{\gamma} = \int d^4 x \sqrt{-g} \delta \left( R_{\mu\nu\alpha\beta}  \ln \left( \dfrac{\Box}{\mu^2} \right) R^{\mu\nu\alpha\beta} \right) \,.
\end{equation}
We see that the variation $\delta S_1$ \eqref{s1} contributes to the equation of motion
\begin{align}
\gamma g_{\sigma \tau}\bigg[R_{\mu\nu\alpha\beta} \ln \left( \dfrac{\Box}{\mu^2} \right) R^{\mu\nu\alpha\beta}\bigg] \,.
\end{align}
But using the same reasoning as \eqref{gaussbonnet} to \eqref{RRterm}, that is, restricting our attention to up to second order curvature corrections, this vanishes when evaluated on the Kerr-AdS solution.

For the variation $\delta S_2$, we use the nonlocal Gauss-Bonnet identity \eqref{gaussbonnet}. By restricting our attention to second order curvature, we may also safely ignore the variation of $\delta \ln \left( \Box / \mu^2 \right)$, which was shown to lead to corrections beyond the quadratic curvature, and commute the curvature tensors safely with $\ln\left(\Box/\mu^2\right)$ as argued in \cite{Calmet:2018elv}. Taking in these considerations, we write down
\begin{align}
\label{vary1}
\delta \bigg[ R_{\mu\nu\alpha\beta}  \ln \left( \dfrac{\Box}{\mu^2} \right) R^{\mu\nu\alpha\beta}\bigg] = \ & 4 \delta R_{\mu\nu} \ln \left(\dfrac{\Box}{\mu^2}\right) R^{\mu\nu}- \delta R \ln \left(\dfrac{\Box}{\mu^2}\right) R \nonumber \\
&  \ \ + \ln \left(\dfrac{\Box}{\mu^2}\right)\bigg( 4 R^{\mu\nu}\delta R_{\mu\nu} - R \delta R \bigg)+ {\cal O}\left(R^3\right).
\end{align}
We also note that
\begin{equation}
    \delta R = \delta g^{\mu\nu} R_{\mu\nu} + g^{\mu\nu} \delta R_{\mu\nu} \,.
\end{equation}
We make further progress by noting that $S_2 \sim \gamma$. Corrections to the geometry thus enter as higher curvature orders in $\delta S_2$, except for the leading AdS metric, i.e., \eqref{eq:ss_ansatz} and \eqref{ef1}. This lets us input
\begin{equation}
    4 R^{\mu\nu}\delta R_{\mu\nu} - R \delta R = -\dfrac{36}{L^4} g_{\mu\nu} \delta g^{\mu\nu} + O \left( L^{-6} \right) \,
\end{equation}
into the variation $\delta S_2$. We are then able to write down
\begin{equation}
\begin{split}
\label{boundary1}
    \delta \left[ R_{\mu\nu\alpha\beta}  \ln \left( \dfrac{\Box}{\mu^2} \right) R^{\mu\nu\alpha\beta} \right]
    = & - \dfrac{36}{L^4} \bigg[ \delta g^{\mu\nu} \ln \left( \dfrac{\Box}{\mu^2} \right) g_{\mu\nu} + g_{\mu\nu} \ln \left( \dfrac{\Box}{\mu^2} \right) \delta g^{\mu\nu} \bigg] \,.
\end{split}
\end{equation}
The first term of the right-hand side of \eqref{boundary1} contributes to the bulk equation of motion of the form
\begin{align}
\label{nonloaceom}
\gamma\frac{36}{L^{{4}}}\ln  \left( \dfrac{\Box}{\mu^2} \right) g_{\mu\nu}=\gamma\frac{36}{L^{{4}}}g_{\mu\nu}\ln\frac{m^2_\text{IR}}{\mu^2+m^2_\text{IR}} \,,
\end{align}
where we used the integral representation \eqref{integralrep} and the infrared regulator $m^2_\text{IR}$. The second term produces a boundary term which we will deal with in Section \ref{boundary}.

Substituting \eqref{nonloaceom} to the equation of motion in the bulk, we see that the nonlocal term seemingly modifies the cosmological constant factor
\begin{align}
\label{lambdaprime}
\Lambda\rightarrow\Lambda'= \ & \Lambda-\gamma\frac{36}{L^{{4}}}\ln\frac{m^2_\text{IR}}{\mu^2+m^2_\text{IR}}\nonumber\\
= \ &\Lambda \left(1+12 \dfrac{\gamma}{{L^2}} \ln\frac{m^2_\text{IR}}{\mu^2+m^2_\text{IR}}\right) \,.
\end{align}
The running of the cosmological constant however is inconsistent with diffeomorphism invariance. The factor in front of $\Lambda$ in \eqref{lambdaprime} can in fact be removed by rescaling the metric \cite{Hamber2013}. After this is done, we end up with the AdS black hole (\eqref{eq:ss_ansatz} and \eqref{ef1}).

Lastly, we note that because the AdS spacetime is not Ricci flat, we need to consider the nonlocal contribution
\begin{align}
\label{eq:nonlocalricci}
S'_\text{nl}=-\alpha \int d^4 x \sqrt{-g} R\ln \frac{\Box}{\mu^2} R \,.
\end{align}
There is no need to consider the action proportional to $R_{\mu\nu}\ln(\Box/\mu^2)R^{\mu\nu}$ since this can be eliminated using the nonlocal Gauss-Bonnet identity \cite{Calmet:2018elv, Calmet:2021lny}. Using similar manipulations we did in \eqref{vary1} to \eqref{boundary1}, the variation of this action gives
\begin{align}
\label{nonlocalricci}
    \delta S'_\text{nl}= \ -2\Lambda \alpha\int d^4 x \sqrt{-g}& \left[ g_{\mu\nu}\delta g^{\mu\nu}\ln\frac{\Box}{\mu^2}R +Rg_{\mu\nu}\ln\frac{\Box}{\mu^2}\delta g^{\mu\nu}\right] \,.
\end{align}
The second line contributes only to the boundary term. The first term contributes to the bulk equation of motion given by
\begin{align}
\label{eomricci}
    L_{\mu\nu}=&-2\Lambda\alpha g_{\mu\nu}\ln\frac{\Box}{\mu^2}R \nonumber \\
    =& \ 16\Lambda^2\alpha g_{\mu\nu}\left[\ln(\mu r)+\gamma_E-1\right] \,.
\end{align}
The contribution of this nonlocal term is therefore of the order
\begin{align}
    \Lambda^2\alpha\sim \frac{\alpha}{L^4} \,,
\end{align}
which is suppressed by the AdS scale compared to those coming from \eqref{nonlocalaction}, whose contributions are of the order ${{\cal O}\left(L^2\right)}$. This is expected on grounds that in the asymptotically flat limit ($L \rightarrow \infty$), the spacetime solution becomes Ricci flat, $R_{\alpha\beta} = 0$, and the terms coming from \eqref{eq:nonlocalricci} vanish while the ones from \eqref{nonlocalaction} make a finite impact \cite{Calmet:2021lny}. In the limit of large \textit{but} finite AdS lengths, the terms coming out of \eqref{eq:nonlocalricci} should therefore be subdominant compared to the ones emerging from \eqref{nonlocalaction}. This is simply reflected by our calculation above.

{Briefly put, the term, $\gamma / L^4 = {\cal O}\left( L^{-2}\right)$ since $\gamma = {\cal O}\left( L^{2}\right)$ if the EFT is to be relevant. On the other hand, the constant $\alpha = {\cal O}\left( L^{0}\right)$ by the physical argument raised in the previous paragraph on the approach to the asymptotically flat case. This, in turn, makes the contribution, $\alpha/L^4$ in (41-42) of ${\cal O}\left(L^{-4}\right)$, which is rightfully subdominant in the large $L$ limit. All boils down to the realization that $\gamma R_{\mu \nu \alpha \beta} \log \left( \Box / \mu^2 \right) R^{\mu\nu \alpha\beta}$ or \eqref{nonlocalaction} remains the significant source of quantum corrections even away from the asymptotically flat limit \cite{Calmet:2021lny}.}

To summarize, the quadratic curvature EFT and nonlocal actions do not give corrections to the AdS metric (\eqref{eq:ss_ansatz} and \eqref{ef1}). This is an important result that logically anchors one of our main results (to be fleshed out Section \ref{thermo}): that the notion of the quantum pressure holds beyond asymptotic flatness.

On the other hand, the cubic curvature terms change the metric as we proceed to show next.

\subsection{Cubic curvature perturbation}
\label{sec:}

The variation of the cubic curvature action \eqref{eq:O3term} leads to the following terms in the field equations
\begin{equation}
\label{eq:eomcubic}
\begin{split}
    \lambda_3 \bigg( & -6 G^{\alpha \delta} R_{\beta \alpha}^{\ \ \mu\nu}R_{\gamma\delta\mu\nu} - 3 R R_{\beta}^{\ \alpha\mu\nu}R_{\gamma\alpha\mu\nu} 
    + 3 R_{\alpha \mu\nu\rho} R_{\beta}^{\ \alpha \delta \mu} R_{\gamma\delta}^{\ \ \nu\rho} \\
    & \ \  + 12 R_{\alpha \nu \mu\rho} R_\beta^{\ \alpha \delta \mu} R_{\gamma \ \delta \ }^{\ \nu \ \rho} + 3 R_{\beta}^{\alpha \delta \mu} R_{\gamma \alpha}^{\ \ \nu\rho} R_{\delta \mu\nu\rho} + 3 R_\beta^{\alpha \delta\mu}R_{\gamma \ \alpha \ }^{\ \nu \ \rho}R_{\delta \mu\nu\rho} \\
    & \ \  -\frac{g_{\beta\gamma}}{2} R_{\alpha\delta}^{\ \ \rho\sigma} R^{\alpha\delta\mu\nu} R_{\mu\nu\rho\sigma} + 6 \nabla_\alpha R_\beta^{\ \alpha\delta\mu} \nabla_\nu R_{\gamma \ \delta \mu}^{\ \nu} \\
    & \ \ + 6 R_\gamma^{\ \alpha\delta\mu} \nabla_\nu \nabla_\alpha R_{\beta \ \delta\mu}^{\ \nu} + 6 R_\beta^{\ \alpha \delta\mu} \nabla_\nu \nabla_\alpha R_{\gamma \ \delta \mu}^{\ \nu} + 6 \nabla_\alpha R_{\gamma \nu\delta\mu} \nabla^\nu R_\beta^{\ \alpha \delta\mu} \bigg) \,.
\end{split}
\end{equation}
To see that this indeed sources the leading order metric correction, we evaluate it on the AdS solution (\eqref{eq:ss_ansatz} and \eqref{ef1}), eventually getting to the nonvanishing contribution, $\propto 4\lambda_3 \Lambda^3 g_{\beta\gamma}/9$. This implies that the cubic curvature EFT anchors a correction to the AdS solution (\eqref{eq:ss_ansatz} and \eqref{ef1}), which we may expect given the analogous results in the asymptotically flat case \cite{Calmet:2021lny}. We now find this metric correction.

We resort to small corrections about the AdS solution \eqref{eq:ss_ansatz} and \eqref{ef1} by treating $\lambda_3$ as a perturbation parameter. To do so, we re-express the Einstein-Hilbert action together with $\Lambda$ and the third order curvature term as
\begin{equation}
\label{eq:metric_O3}
\begin{split}
    ds^2 = & \frac{r^2}{L^2}\left\{- \left[ f_\text{AdS}(r) + \lambda_3 \delta h(r) \right]dt^2 + L^2 d\Omega^2 \right\} +\frac{L^2}{r^2 \left[ f_\text{AdS}(r) + \lambda_3 \delta f(r) \right]}dr^2 \,,
\end{split}
\end{equation}
and then vary the action with respect to $\delta h$ and $\delta f$ to obtain the metric correction at ${\cal O}\left(\lambda_3\right)$. This way, we get to
\begin{equation}
\begin{split}
    \delta h(r) = \ & \frac{4}{\delta L^4 \kappa } + \dfrac{4 \left(L^4-\delta L^4\right)}{\kappa \delta L^4 L^2 r^2} \\
    & \ \ + \dfrac{ 4 M^3 \left(L^{-4}-\delta L^{-4}\right) - \delta M^3 L^{-4}}{ \kappa r^3}
    +\frac{5 M^9}{\kappa  L^4 r^9}+\frac{12 M^6}{\kappa  L^4 r^6} 
\end{split}
\end{equation}
and
\begin{equation}
    \delta f(r) = \frac{4}{\kappa  L^4} -\frac{\delta M^3}{\kappa L^4 r^3} +\frac{66 M^6}{\kappa  L^4 r^6}+\frac{54 M^6}{\kappa  L^2 r^8} -\frac{49 M^9}{\kappa  L^4 r^9} \,,
\end{equation}
where $\delta M$ and $\delta L$ are integration constants. Clearly, however, the constants $\delta M$ and $\delta L$ can be absorbed into a redefinition of the background mass scales $M$ and $L$. This is effectively taken care with the choice $\delta M = 0$ and $\delta L = L$ for which the metric correction is determined by
\begin{equation}
\label{eq:deltah}
\begin{split}
    \delta h(r) = \frac{4}{\kappa L^4} +\frac{5 M^9}{\kappa  L^4 r^9}+\frac{12 M^6}{\kappa  L^4 r^6} 
\end{split}
\end{equation}
and
\begin{equation}
\label{eq:deltaf}
    \delta f(r) = \frac{4}{\kappa  L^4} +\frac{66 M^6}{\kappa  L^4 r^6}+\frac{54 M^6}{\kappa  L^2 r^8} -\frac{49 M^9}{\kappa  L^4 r^9} \,.
\end{equation}

We confirm that these are the generalization of the corresponding third-order curvature EFT corrected Schwarzschild solution presented in \cite{Calmet:2021lny}. This is straightforwardly shown by writing down the metric in the form
\begin{equation}
\label{eq:metricO3}
    ds^2 = - H(r) dt^2 + \dfrac{dr^2}{F(r)} + r^2 d\Omega^2 \,.
\end{equation}
By using \eqref{eq:deltah}, \eqref{eq:deltaf}, and \eqref{eq:Mschw}, we identify
\begin{equation}
\label{eq:metricHO3}
\begin{split}
    H(r) = 1 -\dfrac{2 \mathcal{M}}{r} & + \dfrac{r^2}{L^2} \left(1 + \dfrac{4 \lambda_3 }{\kappa L^4} \right)
    + \frac{48 \mathcal{M}^2 \lambda_3 }{\kappa  L^2 r^4} +\frac{40 \mathcal{M}^3 \lambda_3 }{\kappa  r^7} 
\end{split}
\end{equation}
and
\begin{equation}
\label{eq:metricFO3}
\begin{split}
    F(r) = 1 &- \dfrac{2 \mathcal{M}}{r}+ \dfrac{r^2}{L^2} \left(1 + \frac{4 \lambda_3 }{\kappa  L^4} \right) 
    + \frac{264 \mathcal{M}^2 \lambda_3 }{\kappa  L^2 r^4} +\frac{216 \mathcal{M}^2 \lambda_3 }{\kappa  r^6} -\frac{392 \mathcal{M}^3 \lambda_3 }{\kappa  r^7} \,.
\end{split}
\end{equation}
Therefore, in the asymptotically flat limit, $L \rightarrow \infty$, we have
\begin{equation}
\begin{split}
    H(r) = 1 -\dfrac{2 \mathcal{M}}{r} +\dfrac{40 \mathcal{M}^3 \lambda_3 }{\kappa  r^7} 
\end{split}
\end{equation}
and
\begin{equation}
\begin{split}
    F(r) = 1 &- \dfrac{2 \mathcal{M}}{r} + \dfrac{8 \mathcal{M}^2 \lambda_3}{\kappa  r^6} \left(27 - 49 \dfrac{ \mathcal{M} }{r} \right) \,,
\end{split}
\end{equation}
which completely agrees with \cite{Calmet:2021lny}. The correction ($\sim \lambda_3$) shifts the event horizon by an amount $\delta r = -5 \lambda_3/\left( 8 \kappa  \mathcal{M}^3 \right)$. The above asymptotically flat metric functions and event horizon shift coincide exactly with \cite{Calmet:2021lny} and so supports \eqref{eq:deltah} and \eqref{eq:deltaf} as its proper cubic curvature EFT generalization.

To study the black hole thermal features such as the temperature and the entropy (Section \ref{thermo}), often relying on the event horizon, it is also useful to express the AdS metric in terms of the unperturbed event horizon radius $R_\text{AdS}$, defined as the real solution to
\begin{equation}
    1 + \dfrac{L^2}{R_\text{AdS}^2} - \dfrac{M^3}{R_\text{AdS}^3} = 0 \,.
\end{equation}
The cubic curvature metric functions can be written as
\begin{equation}
\label{eq:HAdS_LR}
\begin{split}
    H\left(r\right) = 1 & +\dfrac{L^2}{r^2} - \dfrac{R_\text{AdS}}{r^3}\left(L^2+R_\text{AdS}^2\right) \\
    & + \dfrac{\lambda_3}{\kappa L^4} \bigg( 4 +\dfrac{12 R_\text{AdS}^2}{r^6} \left(L^2+R_\text{AdS}^2\right)^2 
    +\dfrac{5 R_\text{AdS}^3 }{r^9}\left(L^2+R_\text{AdS}^2\right)^3 \bigg) \,
\end{split}
\end{equation}
and
\begin{equation}
\label{eq:FAdS_LR}
\begin{split}
    F\left(r\right) = 1 & +\dfrac{L^2}{r^2} - \dfrac{R_\text{AdS}}{r^3}\left(L^2+R_\text{AdS}^2\right) \\
    & + \dfrac{ \lambda_3 }{\kappa L^4} \bigg( 4 + \dfrac{66 R_\text{AdS}^2 }{r^6}\left(L^2+R_\text{AdS}^2\right)^2 + \dfrac{54 R_\text{AdS}^2 L^2}{ r^8}\left(L^2+R_\text{AdS}^2\right)^2 \\
    & \phantom{GGGGGg} - \dfrac{49 R_\text{AdS}^3}{r^9}\left(L^2+R_\text{AdS}^2\right)^3 \bigg) \,.
\end{split}
\end{equation}
The event horizon correction $\delta R_\text{AdS}$ can then be shown to be given by
\begin{equation}
\label{eq:deltaRAdS}
    -\dfrac{\kappa L^4 \delta R_\text{AdS}}{\lambda_3} = \dfrac{5 L^6+27 L^4 R_\text{AdS}^2+39 L^2 R_\text{AdS}^4+21 R_\text{AdS}^6}{ L^2 R_\text{AdS}^3 + 3 R_\text{AdS}^5} \,.
\end{equation}
At leading order in the large AdS length $L$ limit, this gives
\begin{equation}
\label{eq:deltaRNearSchw}
    \delta R_\text{AdS} = - \dfrac{5 \lambda_3}{\kappa R_\text{AdS}^3} \left( 1 + \dfrac{12}{5} \dfrac{R_\text{AdS}^2}{ L^2} \right) + {\cal O}\left( \left(R_\text{AdS}/L\right)^3 \right) \,.
\end{equation}
The first term besides $\lambda_3$ in the parenthesis is merely the asymptotically flat correction \cite{Calmet:2021lny}. The second term, ${\cal O}\left(L^{-2}\right)$, is the event horizon correction due to an AdS boundary in the presence of cubic curvature. This shows that the existence of an AdS boundary inevitably enhances the horizon shift anchored on the cubic curvature.

\begin{figure}[h!]
\center
	\subfigure[  ]{
		\includegraphics[width = 0.7 \textwidth]{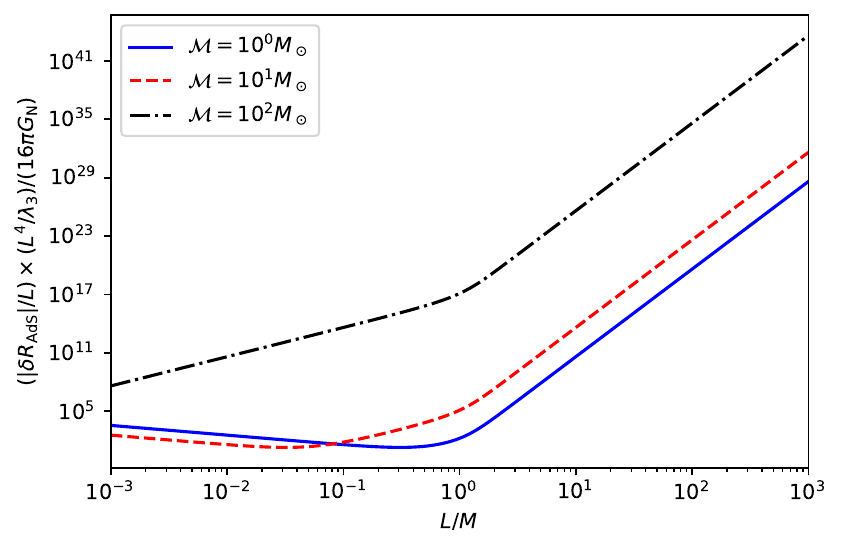}
		}
	\subfigure[  ]{
		\includegraphics[width = 0.7 \textwidth]{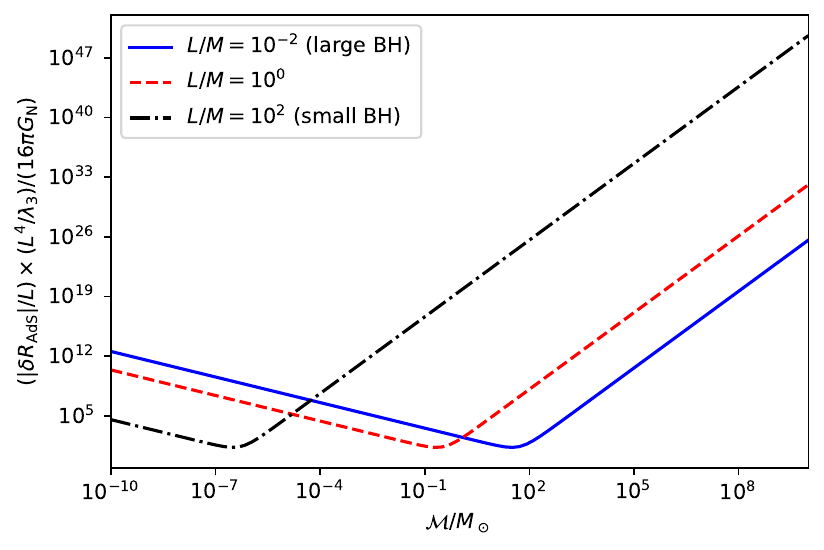}
		}
\caption{AdS black hole event horizon radius correction from cubic curvature EFT, $\delta R_{\rm AdS}$, expressed in the dimensionless $-(\delta R_{\rm AdS}/L)\times(L^4/\lambda_3)\times(1/15\pi G_{\rm N})$, dependencies on (a) $L/M$ and (b) $\cal M$.}
\label{fig:AdShorizonshift}
\end{figure}

Figure \ref{fig:AdShorizonshift} shows explicitly how the position of the event horizon, expressed in terms of the dimensionless quantity, $\left(\delta R_{\rm AdS}/L\right)\times\left(L^4/\lambda_3\right)\times\left(1/16\pi G_{\rm N}\right)$, depend on the mass ratio $L/M$ and the mass $\cal M$.

For a fixed mass $\cal M$, in the small black hole limit, $L/M \gg 1$, we see that the event horizon shift approaches a constant. We understand this in Figure \ref{fig:AdShorizonshift}(a) by expressing the vertical axis as $\delta R_{\rm AdS} (L/M)^9 {\cal M}$. The steep $(L/M)^9$ growth in the small black hole limit thus corresponds to a constant horizon shift, as one would expect from a Schwarzschild black hole, unresponsive to the presence of a boundary at asymptotic infinity. The transition to the large black hole, $L/M = {\cal O}(1)$, is additionally shown in Figure \ref{fig:AdShorizonshift}(a). Once in the large black hole limit, $L/M \ll 1$, we see that the factor $(L/M)^9$ is being suppressed by a growth in the horizon shift $\delta R_{\rm AdS}$, which depends on the mass. Understandably, the shift is larger for the larger mass $\cal M$, but interestingly the rate of increase with decreasing $L/M$ is faster with the smaller mass $\cal M$.

With a fixed ratio $L/M$ determining whether the black hole is small, intermediate, or large, we find the mass $\cal M$ dependence to be more consistent with each other. Recalling that the vertical axis for fixed $L/M$ behaves as ${\cal M}^3 \delta R_{\rm AdS}$, we see that the horizon shift becomes a constant for large $\cal M$, as reflected in the $L/M \gg 1$ region in Figure \ref{fig:AdShorizonshift}(b) where the vertical behavior grows as ${\cal M}^3$. The opposite transition to small masses $\cal M$ is interesting as it occurs in all cases. Figure \ref{fig:AdShorizonshift}(b) shows that at some point, the horizon shift $\delta R_{\rm AdS}$ overcomes the suppressing factor ${\cal M}^3$ for small masses. However, for the small black hole, where $\cal M$ takes the meaning of the Schwarzschild mass, we see that this occurs for only astrophysically irrelevant scales below $\lesssim 10^{-7} M_\odot$. This cubic curvature EFT correction is nonetheless an interesting behavior to further explore.

We also acknowledge that there are nonlocal corrections that settle in to third order in the curvature which are expected to factor in as a scale dependence in $\lambda_3$ \cite{Calmet:2021lny}.

\section{Boundary term}
\label{boundary}
In general relativity calculations, one is usually interested in solving the equations of motion in the bulk. The Gibbons-Hawking-York (GHY) boundary term only serves to cancel the boundary term produced during the variation of the bulk action, intending to make the overall variational principle well posed \cite{Dyer:2008hb}. In AdS/CFT, however, the boundary term plays an important role, where it is dual to the free energy of the dual field theory \cite{Zaanen2015}. Here we will derive the appropriate boundary term corresponding to the nonlocal action but leave working out the full dictionary for future work.

We start by recalling the variation of the bulk nonlocal action \eqref{nonlocalaction} with respect to the metric \eqref{s1ands2}. Only the second part of \eqref{boundary1} leads to a boundary term. For this, we use the expansion
\begin{align}
\label{lnexpand}
    \ln z=\sum_{k=1}^\infty\frac{(-1)^{k+1}}{k}(z-1)^k \,.
\end{align}
Note that this is an expansion about $z=1$ with radius of convergence unity $\mid~z-1\mid<1$. We will replace $z$ with the operator $\Box/\mu^2$, which scales as $\sim L^{-2}/\mu^2$. Hence the expansion point for the logarithm is $L^{-2}\sim\mu^2$. The case of physical interest, $L^{-2} \ll \mu^2$, in which the relevant scales lie further below the EFT cutoff, is covered by the series \eqref{lnexpand} since it is inside the radius of convergence.

The relevant part of the variation of $S_2$ is therefore
\begin{align}
    \delta S_2= \ &\gamma\frac{36}{L^4}\int d^4x\sqrt{-g} g_{\mu\nu}\ln\frac{\Box}{\mu^2}\delta g^{\mu\nu}\nonumber\\
    = \ &\gamma\frac{36}{L^4}\int d^4x\sqrt{-g} g_{\mu\nu}\sum_{k=1}^\infty\frac{(-1)^{k+1}}{k}\left(\frac{\Box}{\mu^2}-1\right)^k\delta g^{\mu\nu}\nonumber\\
    = \ &\gamma\frac{36}{L^4}\int d^4x\sqrt{-g} g_{\mu\nu}\sum_{k=1}^\infty\frac{(-1)^{k+1}}{k}\sum_{l=0}^k\binom{k}{l}\left(\frac{\Box}{\mu^2}\right)^l(-1)^{k-l}\delta g^{\mu\nu} \,.
\end{align}
Note that the boundary contribution from the nonlocal Ricci scalar action \eqref{nonlocalricci} has similar form so that we can simply take $\gamma\rightarrow \gamma+4\alpha$ to include the contribution of such term. We can rewrite the integrand as a total derivative by using $g_{\mu\nu}\Box^l\delta g^{\mu\nu}=\nabla_\alpha \left(g_{\mu\nu}\nabla^\alpha\Box^{l-1}\delta g^{\mu\nu} \right)$. Note that this identity holds also for $l=0$ if we interpret $\Box^{-1}$ as the inverse: $\Box\Box^{-1}=1$. In this case, we proceed as
\begin{align}
  \delta S_2&=\sum_{k=1}^\infty\frac{(-1)^{k+1}}{k}\sum_{l=0}^k\binom{k}{l}(-1)^{k-l}\frac{1}{\mu^{2l}}\gamma\frac{36}{L^4}\int d^4x\sqrt{-g} \nabla_\alpha(g_{\mu\nu}\nabla^\alpha\Box^{l-1}\delta g^{\mu\nu}) \nonumber \\
  &=\sum_{k=1}^\infty\frac{(-1)^{k+1}}{k}\sum_{l=0}^k\binom{k}{l}(-1)^{k-l}\frac{1}{\mu^{2l}} \gamma\frac{36}{L^4}\int_{r=r_\text{B}} d^3y\sqrt{|h|} \hat{r}_\alpha(g_{\mu\nu}\nabla^\alpha\Box^{l-1}\delta g^{\mu\nu}) \,,
\end{align}
where the integral in the last line is evaluated at the boundary at $r=r_\text{B}$. Here, $y$ is the coordinate at the boundary, $\hat{r}_\alpha$ is a unit normal to the boundary, and $h_{\mu\nu}$ is the induced metric. 

We can write $\Box^{l-1}=\Box^{-1}\Box^l$ and bring the summations inside
\begin{align}
    \delta S_2=\ &\gamma\frac{36}{L^4}\int_{r_B} d^3y\sqrt{|h|} \hat{r}_\alpha g_{\mu\nu}\nabla^\alpha\Box^{-1}\sum_{k=1}^\infty\frac{(-1)^{k+1}}{k}\sum_{l=0}^k\binom{k}{l}(-1)^{k-l}\frac{\Box^l}{\mu^{2l}}\delta g^{\mu\nu} \,.
\end{align}
The summations can be evaluated exactly yielding
\begin{align}
\label{boundary2}
     \delta S_2&=\gamma\frac{36}{L^4}\int_{r_\text{B}} d^3y\sqrt{|h|} \hat{r}_\alpha g_{\mu\nu}\nabla^\alpha\Box^{-1}\ln\frac{\Box}{\mu^2}\delta g^{\mu\nu} \,.
\end{align}

We now seek an appropriate nonlocal GHY action that will cancel the boundary term \eqref{boundary2}. From the form of \eqref{boundary2}, a reasonable guess is
\begin{align}
\label{GHY}
    S_\text{GHY}&=\gamma\int_{r_\text{B}} d^3y\sqrt{|h|} R_{\mu\nu\rho\sigma}\hat{r}_\alpha\nabla^\alpha\Box^{-1}\ln\frac{\Box}{\mu^2} R^{\mu\nu\rho\sigma} \,.
\end{align}
Here, the Riemann curvature and the covariant derivatives are to be evaluated first using the bulk metric, then the integrand is evaluated at the boundary $r=r_\text{B}$. Alternatively, we can express the GHY action in terms of the quantities that are intrinsic to the boundary surface. This can be done by foliating the boundary region with fixed-$r$ surfaces then using the ADM formalism. The following formulas that relate the bulk quantities to the surface quantities \cite{Corichi2022} can then be used:
\begin{align}
^4\nabla_{\mathbf{e}_\mu} \mathbf{V}&=^3\nabla_{\mathbf{e}_\mu} \mathbf{V}+^4\Gamma^r_{\mu\lambda}\mathbf{e}_rV^\lambda\\
^4R^d_{\;cab}&=^3R^d_{\;cab}+K_{cb}K^d_{\;a}-K_{ca}K^d_{\;b},\\
R^r_{\nu\alpha\beta}&=\eta_{\nu;\alpha\beta}-\eta_{\nu;\beta\alpha},
\end{align}
where $K_{\mu\nu}$ is the extrinsic curvature and $\eta_\mu$ are the components of unit vector $\hat{r}$ normal to the boundary surface. The connection for the covariant derivative $^3\nabla$ on the boundary surface is $^3\Gamma^c_{ab}=^4\Gamma^c_{ab}$, where the $a$, $b$, and $c$ indices excluding the $r$ coordinate. While it is usually desirable to expressed the boundary terms in terms of the quantities intrinsic to the surface, in this case, it only complicates the final formula without any additional advantage so we leave it in the form \eqref{GHY}.

We now show that this is {a viable candidate for} the desired nonlocal GHY action whose variation cancels the boundary term \eqref{boundary2}. Using the same approximations \cite{Calmet:2018elv} that we used to derived \eqref{vary1} we have
\begin{align}
\label{varyGHY}
\delta S_\text{GHY}= -\gamma\frac{36}{L^4}\int_{r_\text{B}} d^3y\sqrt{|h|} \bigg[ & \delta g^{\mu\nu}\left(\hat{r}_\alpha\nabla^\alpha\Box^{-1}\ln \frac{\Box}{\mu^2}\right)g_{\mu\nu} \nonumber \\
& \ \ +g_{\mu\nu}\left(\hat{r}_\alpha\nabla^\alpha\Box^{-1}\ln \frac{\Box}{\mu^2}\right)\delta g^{\mu\nu} \nonumber \\
& \ \ { + g_{\mu\nu} \left[ \delta \left(\hat{r}_\alpha\nabla^\alpha\Box^{-1} \right) \right] \ln \frac{\Box}{\mu^2} g^{\mu\nu}\bigg] } \,,
\end{align}
 and then eventually get to
\begin{align}
\label{varyGHY2}
\delta S_\text{GHY} =&-\gamma\frac{36}{L^4}\int_{r_\text{B}} d^3y\sqrt{|h|}g_{\mu\nu}\left(\hat{r}_\alpha\nabla^\alpha\Box^{-1}\ln \frac{\Box}{\mu^2}\right)\delta g^{\mu\nu} \nonumber \\
& \ \ { -\gamma\frac{36}{L^4}\int_{r_\text{B}} d^3y\sqrt{|h|}g_{\mu\nu} \left[ \delta \left(\hat{r}_\alpha\nabla^\alpha\Box^{-1} \right) \right] \ln \frac{\Box}{\mu^2} g^{\mu\nu} }
\end{align}
where we used the fact that on the boundary $\delta g^{\mu\nu}=0$ so that the first term inside the square brackets of \eqref{varyGHY} vanishes. Note also that the induced metric is also fixed on this boundary.

{If the contribution due to the variation $\delta \left(\hat{r}_\alpha\nabla^\alpha\Box^{-1}\right)$ vanishes, then} we see that the variation of our proposed nonlocal GHY \eqref{GHY} action cancels the boundary term produced by varying the nonlocal bulk action. {Admittedly, there is a caveat to this that relies on a more grounded definition of the operator $\Box^{-1}$ and its variation, which we leave for future work.}

{We conjecture that equation \eqref{GHY} is the significant piece of a more accurate representation of the boundary term, if there is, that would permit the succinct recasting of overall theory as an action functional.} We can now add this to the known local GHY term \cite{Fukuma2001} to have the full boundary action. {In the following section, we discuss the thermodynamic properties of the theory, extending the established results in the asymptotically flat limit to AdS. We emphasize that this discussion is conservative to the actual form of nonlocal GHY term.}

\section{Thermodynamic quantities}
\label{thermo}
In this section, we outline the calculation of the thermodynamic quantities of the AdS blackhole along with the correction from the higher order curvature terms and the nonlocal action \eqref{nonlocalaction}. We start with the calculation of the Hawking temperature following the prescription of Gibbons and Hawking \cite{Gibbons1993}. 

\subsection{Temperature}
\label{subsec:temperature}
We start with the AdS Schwarzchild metric of the form
\begin{align}
ds^2=\frac{r^2}{L^2}\bigg(-h(r)dt^2+d\Sigma_k^2\bigg)+\frac{L^2}{r^2f(r)}dr^2
\end{align}
where $k=1$ and $k=0$ describe spherical and flat horizons, respectively. Applications of AdS/CFT are mostly concerned with these two possibilities. The spherical case is used when a finite volume dual field theory is needed \cite{Zaanen2015}. Because of this, we will ignore the hyperbolic horizon $k=-1$ case. In the case of a spherical horizon in four dimensions, for concreteness, one may consider the metric \eqref{eq:metric_O3} with the cubic curvature corrections \eqref{eq:deltah} and \eqref{eq:deltaf}, i.e., $d\Sigma_k^2 = L^2 d\Omega^2$ where $d\Omega$ is a differential solid angle. We treat the nonlocal action \eqref{nonlocalaction} and third order curvature terms \eqref{eq:O3term} as weak perturbations.

We are interested with the horizon that is the largest among the solutions $f(r_H)=0$. To calculate the Hawking temperature, we perform the analytic continuation $t\rightarrow -i\tau$ to convert the metric into a Euclidean one
\begin{align}
ds_\text{E}^2=g_{tt}(r)d\tau^2+\frac{dr^2}{g^{rr}(r)}+\frac{r^2}{L^2}d\Sigma_k^2 \,.
\end{align}
We can easily identify Euclidean metric functions with the Minkowski ones as $g_{tt}(r) =r^2 h(r)/L^2$ and $g^{rr}(r) =r^2 f(r)/L^2$. Expanding about the horizon $r=r_H$
\begin{align}
ds_\text{E}^2=g'_{tt}(r_H)(r-r_H)d\tau^2+\frac{dr^2}{g^{rr \prime}(r_H)(r-r_H)}+\frac{r_H^2}{L^2}d\Sigma_k^2
\end{align}
and changing the coordinate
\begin{align}
R_H=\frac{2\sqrt{r-r_H}}{g^{rr \prime}(r_H)},\;\;
\theta=\frac{1}{2}\sqrt{g'_{tt}(r_H)g^{rr \prime}(r_H)}\tau
\end{align}
we get
\begin{align}
ds_\text{E}^2=R_H^2d\theta^2+dR_H^2+\frac{r_H^2}{L^2}d\Sigma_k^2 \,.
\end{align}

Note that $(\theta,R_H)$ looks like a polar coordinate. This analogy is made stronger by requiring that the there should be no singularity at the horizon. This means that the horizon $R_H=0$ should be the origin of the polar coordinate $(\theta, R_H)$. Imposing a periodic boundary on $\theta=\theta+2\pi$ gives a period
\begin{align}
T_\tau=\frac{4\pi}{\sqrt{g'_{tt}(r_H)g^{rr \prime}(r_H)}} \,.
\end{align}
The inverse gives the Hawking temperature
\begin{align}
T=T_\tau^{-1}=\frac{1}{4\pi L^2}\bigg((d+1)r_H+\frac{kL^2(d-1)}{r_H}\bigg) \,,
\end{align}
where $d + 2$ is the number of spacetime dimensions. In four dimensional spacetime ($d = 2$) and up to quadratic curvature and the nonlocal corrections, $\lambda_3 \rightarrow 0$, in which case the metric retains its exact AdS form (\eqref{eq:metric_O3} with $\delta h = \delta f = 0$), the temperature can be shown to be
\begin{equation}
    T = \dfrac{1}{4 \pi L^2} \left( 3r_0 + \dfrac{L^2}{r_0} \right) \,.
\end{equation}
where $r_H = r_0$ is the black hole event horizon. This is the known Hawking temperature of AdS blackhole.

We can separate explicitly the correction $\delta r$ due to the cubic curvature $r_H=r_0+\delta r$ which gives
\begin{align}
\label{eq:hawkingtempcubic}
T = \frac{r_0}{4\pi L^2} & \bigg( ( d+1 ) + \frac{kL^2(d-1)}{r_0^2}\bigg)+ \dfrac{ \delta r }{4\pi L^2} \left( ( d+1 ) - \frac{kL^2(d-1)}{r_0^2} \right) \,.
\end{align}
The second line gives the correction induced by the cubic curvature \eqref{eq:O3term}. With \eqref{eq:deltah} and \eqref{eq:deltaf}, we can show $\delta r$ to be
\begin{equation}
    \delta r = \delta R_\text{AdS}
\end{equation}
where $\delta R_\text{AdS}$ is given by \eqref{eq:deltaRAdS} or, in usual AdS/CFT applications where we are interested in large $L$, the correction $\delta R_\text{AdS}$ is given by \eqref{eq:deltaRNearSchw}. This shows that for flat horizon $k=0$, the correction reduces (increases) the black hole temperature for $\lambda_3>0$ ($\lambda_3 < 0$). The opposite is true for spherical horizon $k=1$. 

\begin{figure}[h!]
\center
	\subfigure[  ]{
		\includegraphics[width = 0.7 \textwidth]{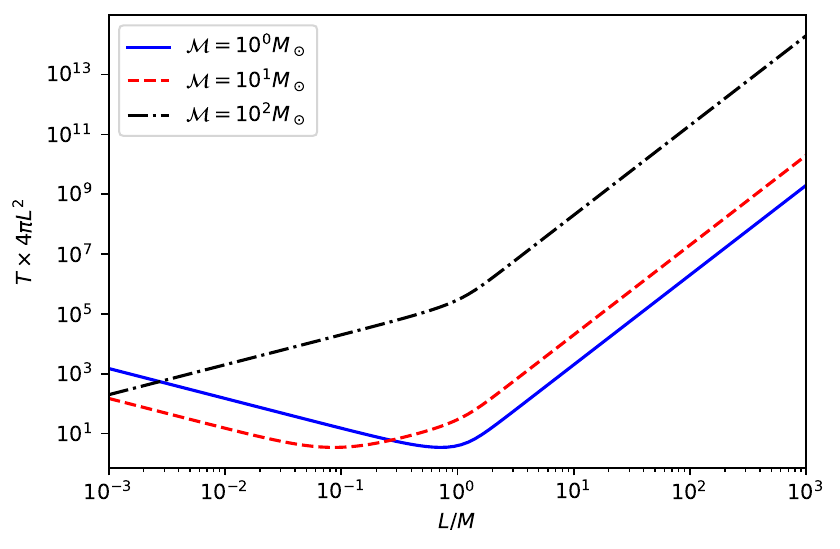}
		}
	\subfigure[  ]{
		\includegraphics[width = 0.7 \textwidth]{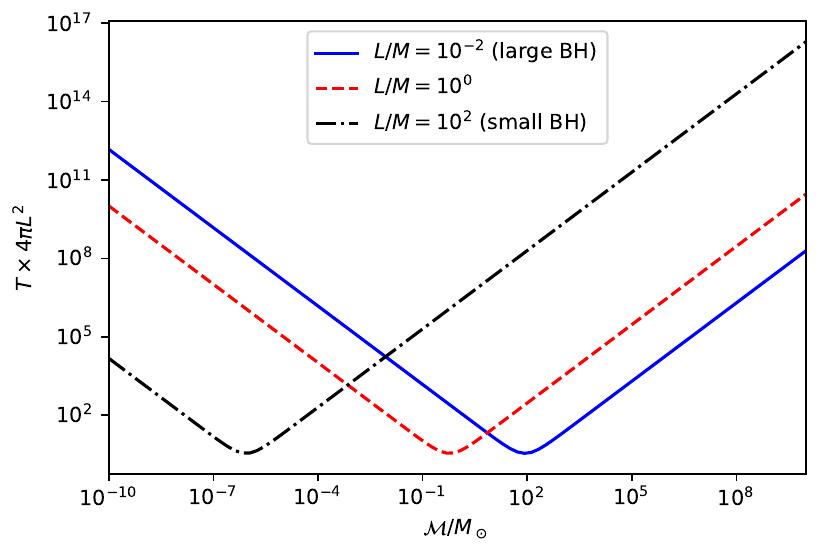}
		}
\caption{AdS black hole temperature $T$, expressed as $T \times \left(4\pi L^2\right)$, as a function of (a) $L/M$ and (b) $\cal M$.}
\label{fig:temp}
\end{figure}

We visualize the temperature profile as a function of $L/M$ and the mass $\cal M$ in Figure \ref{fig:temp}.

This can be understood by rewriting $T L^2$ as 
$T (L/M)^6 {\cal M}^2$. For a fixed $\cal M$ and $T$, we thus expect a scaling of $(L/M)^6$. Instead, we find a smaller scaling with $(L/M)^6$ in the $L/M \gg 1$ region in Figure \ref{fig:temp}(a), revealing the behaviour $T \sim (L/M)^{-3}$. In the large black hole regime, $L/M \ll 1$, we find that the dependence, $T \sim (L/M)^{-7}$, overcoming the suppression factor $(L/M)^{6}$ in Figure \ref{fig:temp}(a). Starting from the large black hole limit, $L/M \ll 1$, we see that the temperature initially decreases as $L/M$ becomes larger. That is, as the black hole becomes smaller. This is consistent with the known behavior of AdS black hole evaporation. If we increase $L/M$ further, it will reach a minimum, a transition point, after which the temperature begins to increase. This happens when the black hole horizon becomes significantly smaller than the AdS radius $L$ so that the black hole becomes effectively a Schwarzschild one. The mass $\cal M$ dependence of the temperature is also shown, where we recognize the familiar inverse mass dependence or the negative black hole heat capacity. In particular, we find the low mass behavior to be $T \sim {\cal M}^{-3}$ and the high mass behavior $T \sim {\cal M}^{-1}$, regardless of the size of the black hole relative to the AdS boundary (Figure \ref{fig:temp}(b)). As a consequence, as a black hole evaporates and losses mass, it heats up faster in an AdS universe. Realizing this with astrophysical black holes however is challenging. As shown in Figure \ref{fig:temp}(b), for a small black hole, the departure $T \sim {\cal M}^{-3}$ from the flat Hawking temperature profile $T \sim {\cal M}^{-1}$ is realized only below $M/M_\odot = {\cal O}(10^{-6})$, outside the known mass ranges of astrophysical black holes in dynamic environments.

\begin{figure}[h!]
\center
	\subfigure[  ]{
		\includegraphics[width = 0.7 \textwidth]{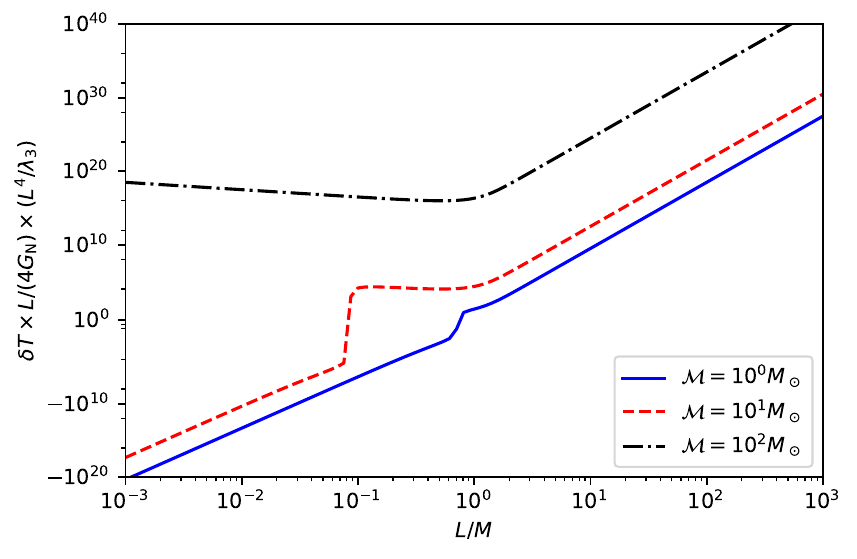}
		}
	\subfigure[  ]{
		\includegraphics[width = 0.7 \textwidth]{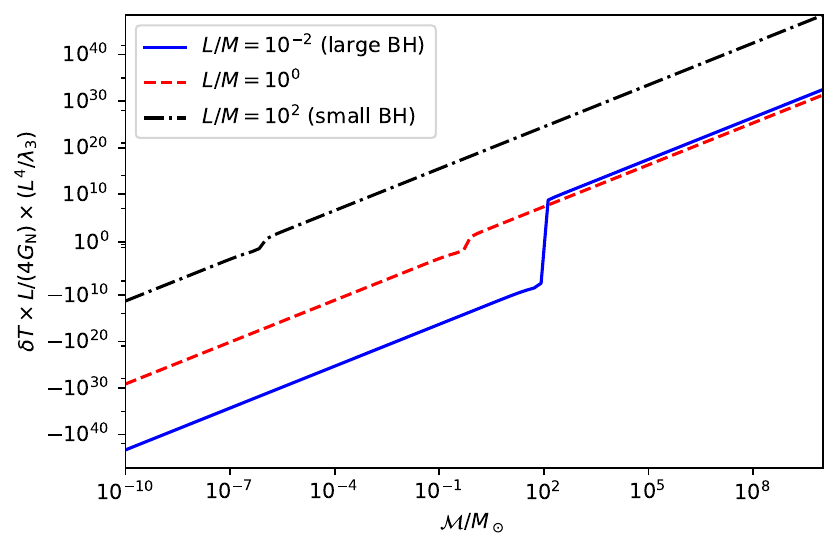}
		}
\caption{AdS black hole temperature shift due to cubic curvature EFT, $\delta T$, expressed as $\delta T \times \left(L/4G_{\rm N}\right) \times (L^4/\lambda_3)$, dependencies on (a) $L/M$ and (b) $\cal M$.}
\label{fig:tempshift}
\end{figure}

We also take a look at the temperature shift brought by the cubic curvature EFT correction (Figure \ref{fig:tempshift}).

First, expressing the vertical behavior in Figure \ref{fig:tempshift} as $\delta T \times (L/M)^{15} {\cal M}^5 \times 8/\lambda_3 G_{\rm N}$, for fixed mass $\cal M$, we realize $\delta T$ approaches a constant in the small black hole limit, $L/M \gg 1$, and the suppression $\delta T \sim (L/M)^{-22}$ in the large black hole limit $L/M \ll 1$. On the mass scaling, we find that the temperature shift approaches a constant for large masses but getting suppressed as $\delta T \sim {\cal M}^{-10}$ for small masses. Interestingly, we interpret our result means that any hope of detecting this temperature shift lie on small black holes where its value becomes a constant. Of course we acknowledge this is wishful thinking since the black holes realized in nature are even colder than the cosmic microwave background ($T \sim 2.7$ kelvins).

\subsection{Entropy}
\label{subsec:entropy}
The addition of the cosmological constant in \eqref{localaction} does not contribute to the variation of the lagrangian density $\delta\mathcal{L}/\delta R_{\mu\nu\rho\sigma}$ in the calculation of the Wald entropy. The entropy then receives corrections due to the presence of an AdS boundary and the nonlocal physics. Also, at third order in the curvature, the metric receives a correction (recall \eqref{eq:metricO3}, \eqref{eq:metricHO3}, \eqref{eq:metricFO3}) and so does the entropy, not only due to the presence of an AdS boundary but also coming from the curvature itself. We calculate these explicitly in this section.

We start with the local terms up to cubic curvature. The relevant action is the sum of \eqref{localaction} and \eqref{eq:O3term}, and the metric is given by \eqref{eq:metric_O3} together with \eqref{eq:deltah} and \eqref{eq:deltaf}. Substituting these into the Wald entropy formula, we obtain
\begin{equation}
\label{eq:entropy_local}
\begin{split}
\mathcal{S}_\text{LW} =
& \dfrac{A}{4G_\text{N}} + 64 \pi^2 c_3(\mu) + \dfrac{128 \pi^3 \lambda_3}{A} - \dfrac{24 \pi  \lambda_3  \left(7 A^2+52 \pi  A L^2+64 \pi ^2 L^4\right)}{3 A L^4+4 \pi  L^6} \,.
\end{split}
\end{equation}
The first line in \eqref{eq:entropy_local} shows the local entropy contributions found for the asymptotically flat black hole \cite{Calmet:2021lny}, while the second line presents the correction due to the presence of the AdS boundary. Clearly, the AdS correction to the black hole entropy is negative, vanishes in the asymptotically flat limit ($L \rightarrow \infty$), and admits the asymptotic expansion
\begin{equation}
\label{eq:entropy_local_asymptotic}
\begin{split}
\mathcal{S}_\text{LW} =
& \dfrac{A}{4G} + 64 \pi^2 c_3(\mu) + \dfrac{128 \pi^3 \lambda_3}{A} -\dfrac{384 \pi ^2 \lambda_3}{L^2}-\dfrac{24 \pi  A \lambda_3}{L^4}+{\cal O}\left(L^{-5}\right) \,.
\end{split}
\end{equation}

Now, the nonlocal correction to the Wald entropy is 
\begin{align}
\mathcal{S}_\text{NLW}=2\pi\gamma\int_{r=r_H}d\Sigma\varepsilon_{\mu\nu}\varepsilon_{\rho\sigma}\ln\left(\frac{\Box}{\mu^2}\right)R^{\mu\nu\rho\sigma},
\end{align}
where $\varepsilon_{tr}=-\varepsilon_{rt}=1$. Thus we only need the Riemann component
\begin{align}
\label{eq:Rtrtr}
R^{trtr} = -\dfrac{M^3}{L^2 r^3} + \dfrac{1}{L^2} \,,
\end{align}
up to antisymmetric permutation of first and last two indices. The actions of the operator $\ln\Box/\mu$ on different radial functions was evaluated in \cite{Delgado2022} while its action on a constant is calculated in \cite{Pourhassan:2022auo}. The first term in \eqref{eq:Rtrtr}, $-2 \mathcal{M}/r^3$ in terms of the Schwarzschild mass \eqref{eq:Mschw}, corresponds to the usual Schwarzschild black hole curvature, while the second term is sourced by the AdS boundary. This gives the finite contribution
\begin{equation}
\label{nls}
\mathcal{S}_\text{NLW} = 16 \pi \gamma a \dfrac{r_H^2}{L^2} \left( \dfrac{M^3}{r_H^3}-\frac{1}{2} \right) \left[ \ln \left( \mu^2 r_H^2 \right) - 2 + 2 \gamma_\text{E} \right] \,,
\end{equation}
where we used the results in \cite{Delgado2022}. Here, $a=4\pi$ for spherical horizon and $a=\int dx^2/ r_H^2$ for flat. The integral of $dx^2$ is infinite, of course, so we should instead speak of entropy density in this case. Since the correction to the metric enters only when the third order curvature is included, the nonlocal entropy correction is sourced mainly by the shift in the event horizon radius induced by the presence of the AdS boundary.

It is instructive to consider the limiting cases. For the Schwarzschild limit $L/M\gg 1$ the horizon is given by $r_H \sim M^3/L^2=2\mathcal{M} $. The entropy correction for spherical case becomes
\begin{align}
\label{entropyschw}
\mathcal{S}_\text{NLW} = \ & 64 \pi^2 \gamma \left[\ln \left(4\mathcal{M}^2\mu^2\right)+2 \gamma_E-2\right]-128\pi^2\gamma\frac{\mathcal{M}^2}{L^2} \left[\ln \left(4\mathcal{M}^2\mu^2\right)+2 \gamma_E-2\right].
\end{align}
This reduces to the nonlocal entropy correction in \eqref{waldentropy2} in the asymptotically flat limit $L\rightarrow\infty$, with $\mathcal{M}$ kept constant.
For large black hole limit $L/M\ll 1$, the horizon is given by \eqref{horizonlarge} and \eqref{nls} yields for the spherical case
\begin{equation}
\label{entropylarge}
\begin{split}
\mathcal{S}_\text{NLW} =\ &32\pi^2\gamma\left(\frac{M^2}{L^2}+\frac{4}{3}\right) 
\left[\ln \left(M^2\mu^2\right)+2 \gamma_E-2\right]\,.
\end{split}
\end{equation}
For flat horizon $k=0$, we have $T\propto r_H$ and the limiting cases above correspond to low and high temperature limits.

\begin{figure}[h!]
\center
	\subfigure[  ]{
		\includegraphics[width = 0.7 \textwidth]{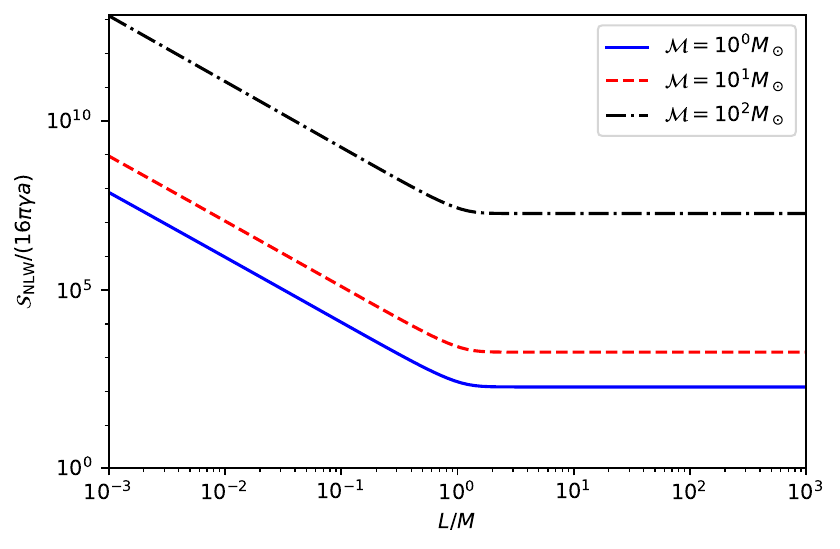}
		}
	\subfigure[  ]{
		\includegraphics[width = 0.7 \textwidth]{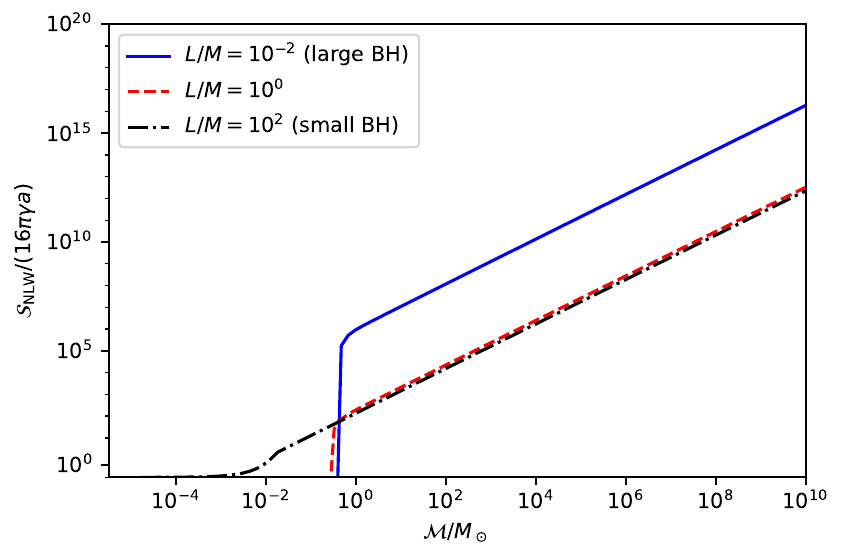}
		}
\caption{AdS black hole quantum entropy \eqref{nls} as a function of (a) $L/M$ and (b) $\cal M$. We keep the cutoff $\mu = 10^{15} \cal M$ fixed throughout.}
\label{fig:quantumentropy}
\end{figure}

The large and small black hole limits of the entropy as well as its mass dependence is further illustrated in Figure \ref{fig:quantumentropy}.

For a fixed mass $\cal M$, the nonlocal entropy approaches a constant $S_{\rm NLW} \sim (L/M)^0$ in the small black hole limit $L/M \gg 1$ but scales as $S_{\rm NLW} \sim (L/M)^{-2}$ in the large black hole limit $L/M \ll 1$ (Figure \ref{fig:quantumentropy}(a)). The entropy thus increase for larger black holes with larger masses. In this case we find the transition between these two regimes to be gradual with the mass ratio $L/M$. The mass dependence with fixed $L/M$ on the other hand looks interesting (Figure \ref{fig:quantumentropy}(b)). For $\cal M \gg M_\odot$ we find that the quantum entropy increases linearly with the mass $S_{\rm NLW} \sim \cal M$. However, in the opposite regime $\cal M \ll M_\odot$ we find an inverse mass scaling, $S_{\rm NLW} \sim -{\cal M}^{-2}$ for ${\cal M} \ll M_\odot$, turning the quantum entropy of very light black holes to be unphysically negative. This holds for small, large, or intermediate size black holes with respect to the AdS boundary. The behavior of the entropy is further interesting during its transition from the high to low mass regime, dictated by the cutoff mass $\mu$. Figure \ref{fig:quantumentropy}(b) shows an abrupt change in the entropy at intermediate masses that is realized when the logarithm in \eqref{nls} turns negative as $\mu r_H \sim 1$. This is more pronounced for large and intermediate black holes. For small black holes, this transition is mild, but can also be realized through a plateau for about three to four orders of masses below the threshold mass when the entropy deviates from the linear mass scaling. Note that the entropy approaches to zero at $\mu r_H \sim 1$. That is, when the horizon approaches the cut-off length scale $\mu^{-1}$ where the system fails to resolve the macroscopic state of the black hole and views it as a single state. Beyond the cut-off $\mu r_H \ll 1$, the theory breaks down and leads to nonphysical negative entropy.

\subsection{Pressure}
\label{subsec:pressure}
Since the horizon does not receive quadratic curvature and leading nonlocal corrections, the entropy correction above gives rise to the quantum pressure \cite{Calmet:2021lny} via the thermodynamic relation $Td\mathcal{S}-PdV$.

For the spherical horizon $k=1$, $V=4\pi r_H^3/3$ and we have the quantum pressures 
\begin{equation}
\begin{split}
P =
& -\frac{12\gamma}{L^4} \left[2 \gamma_E+\ln \left(\mu ^2 M^2\right)-1\right] -\frac{4\gamma}{L^2 M^2} \left[2 \gamma_E+\ln \left(\mu ^2 M^2\right)+3\right] \\
& \ \ -\frac{4\gamma}{3 M^4} \left[6 \gamma_E+3 \ln \left(\mu ^2 M^2\right)+1\right] \,,
\end{split}
\end{equation}
and
\begin{equation}
\label{Schsphere}
P = -\dfrac{\gamma }{2 \mathcal{M}^4}+\dfrac{\gamma}{L^2 \mathcal{M}^2}  \left[2\gamma_E+\ln \left(4 \mu ^2 \mathcal{M}^2\right)-7\right] \,,
\end{equation} \\
for the large black hole $L/M\ll 1$ and Schwarzschild $L/M\gg 1$ limits, respectively. In particular, the Schwarzschild limit \eqref{Schsphere} reduces correctly to the known flat limit ($L\rightarrow\infty$) \cite{Calmet:2021lny} and moreoever presents the leading order quantum correction that arises due to the presence of a boundary. 

For flat horizon $k=0$, we should write the volume as $V=r_Ha$. The quantum pressures in this case are
\begin{equation}
\begin{split}
P =
-\frac{12 \gamma }{L^4}
& +\frac{4\gamma}{L^2 M^2}\left[12 \gamma_E+6 \ln \left(\mu ^2 M^2\right)-17\right] \\
& \ \ +\frac{4\gamma}{3 M^4}\left[32 \gamma_E+16 \ln \left(\mu ^2 M^2\right)-39\right]
\end{split}
\end{equation}
and
\begin{equation}
\begin{split}
P = \ & \frac{\gamma}{2 \mathcal{M}^4} \left[2\gamma_E+\ln \left(4 \mu ^2 \mathcal{M}^2\right)-3\right] + \frac{\gamma}{L^2 \mathcal{M}^2} \left[12 \gamma_E+6 \ln \left(4 \mu ^2 \mathcal{M}^2\right)-17\right]
\end{split}
\end{equation} 
in the large black hole limit and in the Schwarzschild limit, respectively.

\subsection{Free energy}
\label{subsec:freeenergy}
The free energy of the dual field theory is given by the gravitational action evaluated at the bulk metric solution
\begin{align}
    F\equiv -k_\text{B}T\ln \mathcal{Z}_\text{CFT}=k_\text{B}TS^\text{AdS}_\text{E}[g_\text{E}] \,,
\end{align}
where $k_\text{B}$ is the Boltzmann constant. Here the subscript E means that we have analytically continued the metric to the Euclidean one $t\rightarrow i\tau$ with $\tau\in [0,\beta)$, where $\beta=(k_BT)^{-1}$.

We now need to collect all the relevant boundary terms. We have already derived the nonlocal GHY term in \eqref{GHY}. The local GHY terms were derived in \cite{Fukuma2001}. This have the form
\begin{align}
\label{lghy}
    S_\text{lGHY} = \int_0^\beta d\tau\int d^2x\sqrt{h_\text{E}} \bigg( 2K & +x_1RK +x_2R_{ij}K^{ij} \nonumber \\
    & +x_3K^3+x_4KK_{ij}^3+x_5K_{ij}^5 \bigg)
\end{align}
where $K$ is the extrinsic curvature of the boundary at $r=r_\text{B}$. The quantities with lower case Roman indices means they are define on the boundary with induced metric $h_{ij}$. The coefficients $x_i$s are not independent but satisfy certain relations so that the holographic duality holds for higher-derivative gravity as shown in \cite{Fukuma2001}.

The boundary is located at $r_\text{B}\rightarrow\infty$. In this limit the bulk action diverges when the metric solution, which we obtained in Section \ref{ads}, is substituted. This is the UV divergence that can be renormalized by adding an appropriate counter term. We will now show that the appropriate counter term is 
\begin{align}
\label{ct}
 S_\text{ct}=&-\int_0^\beta d\tau\int d^2x\sqrt{h_\text{E}}\frac{2L}{3\sqrt{f(r_\text{B})}} \times 2\Lambda\bigg(\kappa+8c_1\Lambda+c_2\Lambda+\frac{4}{3}c_3\Lambda\bigg) \,.
\end{align}

This can be verified by evaluating the bulk action on-shell using the \eqref{maxsymidentities}. Recall that the nonlocal action just vanishes on-shell as we discussed in Section \ref{ads}. This yields
\begin{align}
    S^\text{o.s.}_\text{EFT}=&\int_0^\beta d\tau\int d^2x\int_{r_0}^{r_\text{B}}dr\frac{r^2}{L^2} \times 2\Lambda\bigg(\kappa+8c_1\Lambda +c_2\Lambda+\frac{4}{3}c_3\Lambda\bigg) \,.
\end{align}

The infrared end at $r_0$ is not relevant to the dual field theory and can be dropped. The ultraviolet end of the boundary at $r_\text{B}$ gives
\begin{align}
    S^\text{o.s.}_\text{EFT}=&\int_0^\beta d\tau\int d^2x\sqrt{h_\text{E}}\frac{2L}{3\sqrt{f(r_\text{B})}} \times 2\Lambda\bigg(\kappa+8c_1\Lambda +c_2\Lambda+\frac{4}{3}c_3\Lambda\bigg) \,.
\end{align}
The factor $\sqrt{h_E}\sim r_B^3$ diverges in the limit $r_B\rightarrow\infty$. This is the ultraviolet divergence of the dual field theory that is precisely canceled by the counter term \eqref{ct}.

The relevant action needed to calculate the free energy is therefore both the nonlocal and local GHY terms \eqref{GHY} and \eqref{lghy} which we collect together
\begin{align}
     S_\text{E}=&\int_0^\beta d\tau\int d^2x\sqrt{h_\text{E}}\bigg(\gamma R_{\mu\nu\rho\sigma}\hat{r}_\alpha\nabla^\alpha\Box^{-1}\ln\frac{\Box}{\mu^2} R^{\mu\nu\rho\sigma} +2K \nonumber \\
     & \phantom{GGGGGGG} +x_1RK+x_2R_{ij}K^{ij}+x_3K^3 +x_4KK_{ij}^3+x_5K_{ij}^5\bigg) \,.
\end{align}

\section{Conclusion}

We have considered the effect of the nonlocal term on the AdS black hole, and found that this produces a quantum pressure that generalizes the well established phenomena for a Schwarzschild black hole \cite{Calmet:2021lny}. In particular, following the logic that lead to the notion of the quantum pressure in a Schwarzschild black hole, we have shown that the quadratic curvature and the non-local actions do not geometrically backreact on the AdS black hole. As a consequence, the nonlocal correction to the entropy (calculated via Wald prescription) leads to the quantum pressure interpretation for AdS black holes. We note that a recent paper obtained second order and nonlocal corrections to the metric using a more specialized variation of the form $g_{\mu\nu}\rightarrow e^\epsilon g_{\mu\nu}$ \cite{Xiao2022}. In contrast, we obtained no correction, which is consistent to the results in \cite{Calmet:2021lny,Calmet2017} for the asymptotically flat black hole. In addition, we calculated the cubic curvature EFT correction to the AdS black hole metric. This is shown to reduce to the known Schwarzschild solution \cite{Calmet:2021lny} in the appropriate limit.

In Section \ref{boundary} we derived the boundary term for the action. Specifically, we conjecture a nonlocal Gibbons-Hawking-York boundary term \eqref{GHY} and showed that part of this cancels the boundary term produced from the metric variation of the bulk nonlocal action. In addition, we also give the appropriate counter term \eqref{ct} that cancels the ultraviolet divergence when the metric solution is substituted to the bulk action and the boundary limit is taken. 

Lastly, we have calculated the thermodynamic quantities such as temperature, entropy, and pressure. We found that the behavior of temperature versus $L/M$ undergoes a transition from decreasing to increasing as the large black hole evaporates towards a smaller black hole. For entropy, we found that it drops to zero as the black hole horizon approaches the cut-off length scale $\mu^{-1}$. We derived the boundary actions needed for the dual free energy. However, the nonlocal Gibbons-Hawking-York boundary action both contains infrared and ultraviolet (from $\sqrt{h}\sim r^3_\text{B}$) divergences. Interestingly, the infrared divergence appears in the boundary dual which is associated with the ultraviolet scale. This is the manifestation of the nonlocality of the quantum gravitational correction. The infrared divergence most likely comes from the self-interacting gravitons and soft gravitons \cite{Tsamis1995}. We leave the investigation of these infinities for future work.

We emphasize that by the correspondence principle our results generalize those for the asymptotically flat black hole \cite{Calmet:2021lny} to the anti-de Sitter black hole. The near Schwarzschild limit which accounts for the leading ${\cal O}\left(L^{-2}\right)$ corrections (e.g., event horizon \eqref{eq:deltaRNearSchw} and quantum pressure \eqref{Schsphere}) thus potentially gives the first astrophysical signals we can expect due to a spacetime boundary. Quasinormal signatures can for example be triggered by the geometrical corrections due to the boundary and cubic curvature effective field theory near the black hole \cite{Cardoso:2018ptl, Cardoso:2019mqo}. Of course these corrections are presumably tiny, if even reachable by any future astrophysical probes. Nonetheless even nondetection can place useful constraints, particularly on the quantum corrections, that can be considered in future work.

\section*{Declarations}


\begin{itemize}
\item Data availability: Data sharing not applicable to this article as no datasets were generated or analysed during the current study.
\item Competing interests: The authors have no competing interests to declare that are relevant to the content of this article.
\item Code availability: Mathematica and python notebooks for the derivation and analysis of the results of this work can be downloaded from \href{https://github.com/reggiebernardo/notebooks}{GitHub} \cite{reggie_bernardo_4810864}.
\end{itemize}

\providecommand{\noopsort}[1]{}\providecommand{\singleletter}[1]{#1}%



\begin{thebibliography}{44}
\ifx \bisbn   \undefined \def \bisbn  #1{ISBN #1}\fi
\ifx \binits  \undefined \def \binits#1{#1}\fi
\ifx \bauthor  \undefined \def \bauthor#1{#1}\fi
\ifx \batitle  \undefined \def \batitle#1{#1}\fi
\ifx \bjtitle  \undefined \def \bjtitle#1{#1}\fi
\ifx \bvolume  \undefined \def \bvolume#1{\textbf{#1}}\fi
\ifx \byear  \undefined \def \byear#1{#1}\fi
\ifx \bissue  \undefined \def \bissue#1{#1}\fi
\ifx \bfpage  \undefined \def \bfpage#1{#1}\fi
\ifx \blpage  \undefined \def \blpage #1{#1}\fi
\ifx \burl  \undefined \def \burl#1{\textsf{#1}}\fi
\ifx \doiurl  \undefined \def \doiurl#1{\url{https://doi.org/#1}}\fi
\ifx \betal  \undefined \def \betal{\textit{et al.}}\fi
\ifx \binstitute  \undefined \def \binstitute#1{#1}\fi
\ifx \binstitutionaled  \undefined \def \binstitutionaled#1{#1}\fi
\ifx \bctitle  \undefined \def \bctitle#1{#1}\fi
\ifx \beditor  \undefined \def \beditor#1{#1}\fi
\ifx \bpublisher  \undefined \def \bpublisher#1{#1}\fi
\ifx \bbtitle  \undefined \def \bbtitle#1{#1}\fi
\ifx \bedition  \undefined \def \bedition#1{#1}\fi
\ifx \bseriesno  \undefined \def \bseriesno#1{#1}\fi
\ifx \blocation  \undefined \def \blocation#1{#1}\fi
\ifx \bsertitle  \undefined \def \bsertitle#1{#1}\fi
\ifx \bsnm \undefined \def \bsnm#1{#1}\fi
\ifx \bsuffix \undefined \def \bsuffix#1{#1}\fi
\ifx \bparticle \undefined \def \bparticle#1{#1}\fi
\ifx \barticle \undefined \def \barticle#1{#1}\fi
\bibcommenthead
\ifx \bconfdate \undefined \def \bconfdate #1{#1}\fi
\ifx \botherref \undefined \def \botherref #1{#1}\fi
\ifx \url \undefined \def \url#1{\textsf{#1}}\fi
\ifx \bchapter \undefined \def \bchapter#1{#1}\fi
\ifx \bbook \undefined \def \bbook#1{#1}\fi
\ifx \bcomment \undefined \def \bcomment#1{#1}\fi
\ifx \oauthor \undefined \def \oauthor#1{#1}\fi
\ifx \citeauthoryear \undefined \def \citeauthoryear#1{#1}\fi
\ifx \endbibitem  \undefined \def \endbibitem {}\fi
\ifx \bconflocation  \undefined \def \bconflocation#1{#1}\fi
\ifx \arxivurl  \undefined \def \arxivurl#1{\textsf{#1}}\fi
\csname PreBibitemsHook\endcsname

\bibitem{Maldacena1998}
\begin{barticle}
\bauthor{\bsnm{Maldacena}, \binits{J.}}:
\batitle{{The Large N Limit of Superconformal field theories and
  supergravity}}.
\bjtitle{Adv. Theor. Math. Phys.}
\bvolume{2},
\bfpage{231}--\blpage{252}
(\byear{1998})
{\href{https://arxiv.org/abs/9711200}{{arXiv:9711200}}}
{[hep-th]}.
\doiurl{10.4310/ATMP.1998.V2.N2.A1}
\end{barticle}
\endbibitem

\bibitem{Kovtun2005}
\begin{barticle}
\bauthor{\bsnm{Kovtun}, \binits{P.}},
\bauthor{\bsnm{Son}, \binits{D.T.}},
\bauthor{\bsnm{Starinets}, \binits{A.O.}}:
\batitle{{Viscosity in strongly interacting quantum field theories from black
  hole physics}}.
\bjtitle{Phys. Rev. Lett.}
\bvolume{94},
\bfpage{111601}
(\byear{2005})
{\href{https://arxiv.org/abs/hep-th/0405231}{{arXiv:hep-th/0405231}}}.
\doiurl{10.1103/PhysRevLett.94.111601}
\end{barticle}
\endbibitem

\bibitem{Heinz2012}
\begin{barticle}
\bauthor{\bsnm{Heinz}, \binits{U.}},
\bauthor{\bsnm{Shen}, \binits{C.}},
\bauthor{\bsnm{Song}, \binits{H.}}:
\batitle{{The viscosity of quark-gluon plasma at RHIC and the LHC}}.
\bjtitle{AIP Conf. Proc.}
\bvolume{1441}(\bissue{1}),
\bfpage{766}--\blpage{770}
(\byear{2012})
{\href{https://arxiv.org/abs/1108.5323}{{arXiv:1108.5323}}}
{[nucl-th]}.
\doiurl{10.1063/1.3700674}
\end{barticle}
\endbibitem

\bibitem{Hartnoll2008}
\begin{barticle}
\bauthor{\bsnm{Hartnoll}, \binits{S.A.}},
\bauthor{\bsnm{Herzog}, \binits{C.P.}},
\bauthor{\bsnm{Horowitz}, \binits{G.T.}}:
\batitle{{Building a Holographic Superconductor}}.
\bjtitle{Phys. Rev. Lett.}
\bvolume{101},
\bfpage{031601}
(\byear{2008})
{\href{https://arxiv.org/abs/0803.3295}{{arXiv:0803.3295}}}
{[hep-th]}.
\doiurl{10.1103/PhysRevLett.101.031601}
\end{barticle}
\endbibitem

\bibitem{Zaanen2015}
\begin{bbook}
\bauthor{\bsnm{Zaanen}, \binits{J.}},
\bauthor{\bsnm{Sun}, \binits{Y.-W.}},
\bauthor{\bsnm{Liu}, \binits{Y.}},
\bauthor{\bsnm{Schalm}, \binits{K.}}:
\bbtitle{{Holographic Duality in Condensed Matter Physics}}.
\bpublisher{Cambridge Univ. Press}, \blocation{???}
(\byear{2015})
\end{bbook}
\endbibitem

\bibitem{Nosaka2020}
\begin{barticle}
\bauthor{\bsnm{Nosaka}, \binits{T.}},
\bauthor{\bsnm{Numasawa}, \binits{T.}}:
\batitle{{Quantum Chaos, Thermodynamics and Black Hole Microstates in the mass
  deformed SYK model}}.
\bjtitle{JHEP}
\bvolume{08},
\bfpage{081}
(\byear{2020})
{\href{https://arxiv.org/abs/1912.12302}{{arXiv:1912.12302}}}
{[hep-th]}.
\doiurl{10.1007/JHEP08(2020)081}
\end{barticle}
\endbibitem

\bibitem{Lensky2021}
\begin{barticle}
\bauthor{\bsnm{Lensky}, \binits{Y.D.}},
\bauthor{\bsnm{Qi}, \binits{X.-L.}}:
\batitle{{Rescuing a black hole in the large-$q$ coupled SYK model}}.
\bjtitle{JHEP}
\bvolume{04},
\bfpage{116}
(\byear{2021})
{\href{https://arxiv.org/abs/2012.15798}{{arXiv:2012.15798}}}
{[hep-th]}.
\doiurl{10.1007/JHEP04(2021)116}
\end{barticle}
\endbibitem

\bibitem{Myers2010}
\begin{barticle}
\bauthor{\bsnm{Myers}, \binits{R.C.}},
\bauthor{\bsnm{Robinson}, \binits{B.}}:
\batitle{{Black Holes in Quasi-topological Gravity}}.
\bjtitle{JHEP}
\bvolume{08},
\bfpage{067}
(\byear{2010})
{\href{https://arxiv.org/abs/1003.5357}{{arXiv:1003.5357}}}
{[gr-qc]}.
\doiurl{10.1007/JHEP08(2010)067}
\end{barticle}
\endbibitem

\bibitem{Parvizi2019}
\begin{barticle}
\bauthor{\bsnm{Parvizi}, \binits{S.}},
\bauthor{\bsnm{Sadeghi}, \binits{M.}}:
\batitle{{Holographic Aspects of a Higher Curvature Massive Gravity}}.
\bjtitle{Eur. Phys. J. C}
\bvolume{79}(\bissue{2}),
\bfpage{113}
(\byear{2019})
{\href{https://arxiv.org/abs/1704.00441}{{arXiv:1704.00441}}}
{[hep-th]}.
\doiurl{10.1140/epjc/s10052-019-6631-9}
\end{barticle}
\endbibitem

\bibitem{Bekenstein1973}
\begin{barticle}
\bauthor{\bsnm{Bekenstein}, \binits{J.D.}}:
\batitle{{Black holes and entropy}}.
\bjtitle{Phys. Rev. D}
\bvolume{7},
\bfpage{2333}--\blpage{2346}
(\byear{1973}).
\doiurl{10.1103/PhysRevD.7.2333}
\end{barticle}
\endbibitem

\bibitem{Hawking1975}
\begin{barticle}
\bauthor{\bsnm{Hawking}, \binits{S.W.}}:
\batitle{{Particle Creation by Black Holes}}.
\bjtitle{Commun. Math. Phys.}
\bvolume{43},
\bfpage{199}--\blpage{220}
(\byear{1975}).
\doiurl{10.1007/BF02345020}.
\bcomment{[Erratum: Commun.Math.Phys. 46, 206 (1976)]}
\end{barticle}
\endbibitem

\bibitem{Wald2001}
\begin{barticle}
\bauthor{\bsnm{Wald}, \binits{R.M.}}:
\batitle{{The thermodynamics of black holes}}.
\bjtitle{Living Rev. Rel.}
\bvolume{4},
\bfpage{6}
(\byear{2001})
{\href{https://arxiv.org/abs/gr-qc/9912119}{{arXiv:gr-qc/9912119}}}.
\doiurl{10.12942/lrr-2001-6}
\end{barticle}
\endbibitem

\bibitem{Kastor:2009wy}
\begin{barticle}
\bauthor{\bsnm{Kastor}, \binits{D.}},
\bauthor{\bsnm{Ray}, \binits{S.}},
\bauthor{\bsnm{Traschen}, \binits{J.}}:
\batitle{{Enthalpy and the Mechanics of AdS Black Holes}}.
\bjtitle{Class. Quant. Grav.}
\bvolume{26},
\bfpage{195011}
(\byear{2009})
{\href{https://arxiv.org/abs/0904.2765}{{arXiv:0904.2765}}}
{[hep-th]}.
\doiurl{10.1088/0264-9381/26/19/195011}
\end{barticle}
\endbibitem

\bibitem{Kastor:2010gq}
\begin{barticle}
\bauthor{\bsnm{Kastor}, \binits{D.}},
\bauthor{\bsnm{Ray}, \binits{S.}},
\bauthor{\bsnm{Traschen}, \binits{J.}}:
\batitle{{Smarr Formula and an Extended First Law for Lovelock Gravity}}.
\bjtitle{Class. Quant. Grav.}
\bvolume{27},
\bfpage{235014}
(\byear{2010})
{\href{https://arxiv.org/abs/1005.5053}{{arXiv:1005.5053}}}
{[hep-th]}.
\doiurl{10.1088/0264-9381/27/23/235014}
\end{barticle}
\endbibitem

\bibitem{Kastor:2011qp}
\begin{barticle}
\bauthor{\bsnm{Kastor}, \binits{D.}},
\bauthor{\bsnm{Ray}, \binits{S.}},
\bauthor{\bsnm{Traschen}, \binits{J.}}:
\batitle{{Mass and Free Energy of Lovelock Black Holes}}.
\bjtitle{Class. Quant. Grav.}
\bvolume{28},
\bfpage{195022}
(\byear{2011})
{\href{https://arxiv.org/abs/1106.2764}{{arXiv:1106.2764}}}
{[hep-th]}.
\doiurl{10.1088/0264-9381/28/19/195022}
\end{barticle}
\endbibitem

\bibitem{Dolan2011a}
\begin{barticle}
\bauthor{\bsnm{Dolan}, \binits{B.P.}}:
\batitle{{The cosmological constant and the black hole equation of state}}.
\bjtitle{Class. Quant. Grav.}
\bvolume{28},
\bfpage{125020}
(\byear{2011})
{\href{https://arxiv.org/abs/1008.5023}{{arXiv:1008.5023}}}
{[gr-qc]}.
\doiurl{10.1088/0264-9381/28/12/125020}
\end{barticle}
\endbibitem

\bibitem{Dolan2011b}
\begin{barticle}
\bauthor{\bsnm{Dolan}, \binits{B.P.}}:
\batitle{{Pressure and volume in the first law of black hole thermodynamics}}.
\bjtitle{Class. Quant. Grav.}
\bvolume{28},
\bfpage{235017}
(\byear{2011})
{\href{https://arxiv.org/abs/1106.6260}{{arXiv:1106.6260}}}
{[gr-qc]}.
\doiurl{10.1088/0264-9381/28/23/235017}
\end{barticle}
\endbibitem

\bibitem{Calmet:2021lny}
\begin{barticle}
\bauthor{\bsnm{Calmet}, \binits{X.}},
\bauthor{\bsnm{Kuipers}, \binits{F.}}:
\batitle{{Quantum gravitational corrections to the entropy of a Schwarzschild
  black hole}}.
\bjtitle{Phys. Rev. D}
\bvolume{104}(\bissue{6}),
\bfpage{066012}
(\byear{2021})
{\href{https://arxiv.org/abs/2108.06824}{{arXiv:2108.06824}}}
{[hep-th]}.
\doiurl{10.1103/PhysRevD.104.066012}
\end{barticle}
\endbibitem

\bibitem{Pourhassan:2022auo}
\begin{botherref}
\oauthor{\bsnm{Pourhassan}, \binits{B.}},
\oauthor{\bsnm{Delgado~Campos}, \binits{R.}}:
{Quantum Gravitational Corrections to the Geometry of Charged AdS Black Holes}
\end{botherref}
\endbibitem

\bibitem{reggie_bernardo_4810864}
\begin{botherref}
\oauthor{\bsnm{Bernardo}, \binits{R.}}:
{reggiebernardo/notebooks: dark energy research notebooks}.
Zenodo
(2021)
\end{botherref}
\endbibitem

\bibitem{Barvinsky1983}
\begin{barticle}
\bauthor{\bsnm{Barvinsky}, \binits{A.O.}},
\bauthor{\bsnm{Vilkovisky}, \binits{G.A.}}:
\batitle{{The generalized Schwinger-DeWitt technique and the unique effective
  action in quantum gravity}}.
\bjtitle{Phys. Lett. B}
\bvolume{131},
\bfpage{313}--\blpage{318}
(\byear{1983}).
\doiurl{10.1016/0370-2693(83)90506-3}
\end{barticle}
\endbibitem

\bibitem{Donoghue1994}
\begin{barticle}
\bauthor{\bsnm{Donoghue}, \binits{J.F.}}:
\batitle{{General relativity as an effective field theory: The leading quantum
  corrections}}.
\bjtitle{Phys. Rev. D}
\bvolume{50},
\bfpage{3874}--\blpage{3888}
(\byear{1994})
{\href{https://arxiv.org/abs/gr-qc/9405057}{{arXiv:gr-qc/9405057}}}.
\doiurl{10.1103/PhysRevD.50.3874}
\end{barticle}
\endbibitem

\bibitem{Calmet2017}
\begin{barticle}
\bauthor{\bsnm{Calmet}, \binits{X.}},
\bauthor{\bsnm{El-Menoufi}, \binits{B.K.}}:
\batitle{{Quantum Corrections to Schwarzschild Black Hole}}.
\bjtitle{Eur. Phys. J. C}
\bvolume{77}(\bissue{4}),
\bfpage{243}
(\byear{2017})
{\href{https://arxiv.org/abs/1704.00261}{{arXiv:1704.00261}}}
{[hep-th]}.
\doiurl{10.1140/epjc/s10052-017-4802-0}
\end{barticle}
\endbibitem

\bibitem{Calmet:2018elv}
\begin{barticle}
\bauthor{\bsnm{Calmet}, \binits{X.}}:
\batitle{{Vanishing of Quantum Gravitational Corrections to Vacuum Solutions of
  General Relativity at Second Order in Curvature}}.
\bjtitle{Phys. Lett. B}
\bvolume{787},
\bfpage{36}--\blpage{38}
(\byear{2018})
{\href{https://arxiv.org/abs/1810.09719}{{arXiv:1810.09719}}}
{[hep-th]}.
\doiurl{10.1016/j.physletb.2018.10.040}
\end{barticle}
\endbibitem

\bibitem{Gubser1998}
\begin{barticle}
\bauthor{\bsnm{Gubser}, \binits{S.S.}},
\bauthor{\bsnm{Klebanov}, \binits{I.R.}},
\bauthor{\bsnm{Polyakov}, \binits{A.M.}}:
\batitle{{Gauge theory correlators from noncritical string theory}}.
\bjtitle{Phys. Lett. B}
\bvolume{428},
\bfpage{105}--\blpage{114}
(\byear{1998})
{\href{https://arxiv.org/abs/hep-th/9802109}{{arXiv:hep-th/9802109}}}.
\doiurl{10.1016/S0370-2693(98)00377-3}
\end{barticle}
\endbibitem

\bibitem{Witten1998}
\begin{barticle}
\bauthor{\bsnm{Witten}, \binits{E.}}:
\batitle{{Anti-de Sitter space and holography}}.
\bjtitle{Adv. Theor. Math. Phys.}
\bvolume{2},
\bfpage{253}--\blpage{291}
(\byear{1998})
{\href{https://arxiv.org/abs/hep-th/9802150}{{arXiv:hep-th/9802150}}}.
\doiurl{10.4310/ATMP.1998.v2.n2.a2}
\end{barticle}
\endbibitem

\bibitem{Tsai:2011gv}
\begin{barticle}
\bauthor{\bsnm{Tsai}, \binits{Y.-D.}},
\bauthor{\bsnm{Wu}, \binits{X.N.}},
\bauthor{\bsnm{Yang}, \binits{Y.}}:
\batitle{{Phase Structure of Kerr-AdS Black Hole}}.
\bjtitle{Phys. Rev. D}
\bvolume{85},
\bfpage{044005}
(\byear{2012})
{\href{https://arxiv.org/abs/1104.0502}{{arXiv:1104.0502}}}
{[hep-th]}.
\doiurl{10.1103/PhysRevD.85.044005}
\end{barticle}
\endbibitem

\bibitem{Charmousis:2019fre}
\begin{barticle}
\bauthor{\bsnm{Charmousis}, \binits{C.}},
\bauthor{\bsnm{Crisostomi}, \binits{M.}},
\bauthor{\bsnm{Langlois}, \binits{D.}},
\bauthor{\bsnm{Noui}, \binits{K.}}:
\batitle{{Perturbations of a rotating black hole in DHOST theories}}.
\bjtitle{Class. Quant. Grav.}
\bvolume{36}(\bissue{23}),
\bfpage{235008}
(\byear{2019})
{\href{https://arxiv.org/abs/1907.02924}{{arXiv:1907.02924}}}
{[gr-qc]}.
\doiurl{10.1088/1361-6382/ab4fb1}
\end{barticle}
\endbibitem

\bibitem{Cardoso:2001bb}
\begin{barticle}
\bauthor{\bsnm{Cardoso}, \binits{V.}},
\bauthor{\bsnm{Lemos}, \binits{J.P.S.}}:
\batitle{{Quasinormal modes of Schwarzschild anti-de Sitter black holes:
  Electromagnetic and gravitational perturbations}}.
\bjtitle{Phys. Rev. D}
\bvolume{64},
\bfpage{084017}
(\byear{2001})
{\href{https://arxiv.org/abs/gr-qc/0105103}{{arXiv:gr-qc/0105103}}}.
\doiurl{10.1103/PhysRevD.64.084017}
\end{barticle}
\endbibitem

\bibitem{Cardoso:2003cj}
\begin{barticle}
\bauthor{\bsnm{Cardoso}, \binits{V.}},
\bauthor{\bsnm{Konoplya}, \binits{R.}},
\bauthor{\bsnm{Lemos}, \binits{J.P.S.}}:
\batitle{{Quasinormal frequencies of Schwarzschild black holes in anti-de
  Sitter space-times: A Complete study on the asymptotic behavior}}.
\bjtitle{Phys. Rev. D}
\bvolume{68},
\bfpage{044024}
(\byear{2003})
{\href{https://arxiv.org/abs/gr-qc/0305037}{{arXiv:gr-qc/0305037}}}.
\doiurl{10.1103/PhysRevD.68.044024}
\end{barticle}
\endbibitem

\bibitem{Wang:2021uix}
\begin{barticle}
\bauthor{\bsnm{Wang}, \binits{M.}},
\bauthor{\bsnm{Chen}, \binits{Z.}},
\bauthor{\bsnm{Pan}, \binits{Q.}},
\bauthor{\bsnm{Jing}, \binits{J.}}:
\batitle{{Maxwell quasinormal modes on a global monopole Schwarzschild-anti-de
  Sitter black hole with Robin boundary conditions}}.
\bjtitle{Eur. Phys. J. C}
\bvolume{81}(\bissue{5}),
\bfpage{469}
(\byear{2021})
{\href{https://arxiv.org/abs/2105.10951}{{arXiv:2105.10951}}}
{[gr-qc]}.
\doiurl{10.1140/epjc/s10052-021-09149-x}
\end{barticle}
\endbibitem

\bibitem{Wang:2021upj}
\begin{barticle}
\bauthor{\bsnm{Wang}, \binits{M.}},
\bauthor{\bsnm{Chen}, \binits{Z.}},
\bauthor{\bsnm{Tong}, \binits{X.}},
\bauthor{\bsnm{Pan}, \binits{Q.}},
\bauthor{\bsnm{Jing}, \binits{J.}}:
\batitle{{Bifurcation of the Maxwell quasinormal spectrum on asymptotically
  anti\textendash{}de Sitter black holes}}.
\bjtitle{Phys. Rev. D}
\bvolume{103}(\bissue{6}),
\bfpage{064079}
(\byear{2021})
{\href{https://arxiv.org/abs/2104.04970}{{arXiv:2104.04970}}}
{[gr-qc]}.
\doiurl{10.1103/PhysRevD.103.064079}
\end{barticle}
\endbibitem

\bibitem{Fortuna:2022fdd}
\begin{barticle}
\bauthor{\bsnm{Fortuna}, \binits{S.}},
\bauthor{\bsnm{Vega}, \binits{I.}}:
\batitle{{Electromagnetic quasinormal modes of Schwarzschild-anti-de Sitter
  black holes: bifurcations, spectral similarity, and exact solutions in the
  large black hole limit}}.
\bjtitle{Phys. Rev. D}
\bvolume{106},
\bfpage{084028}
(\byear{2022})
{\href{https://arxiv.org/abs/gr-qc/0105103}{{arXiv:gr-qc/0105103}}}.
\doiurl{10.1103/PhysRevD.106.084028}
\end{barticle}
\endbibitem

\bibitem{Barvinsky1990}
\begin{barticle}
\bauthor{\bsnm{Barvinsky}, \binits{A.O.}},
\bauthor{\bsnm{Vilkovisky}, \binits{G.A.}}:
\batitle{{Covariant perturbation theory. 2: Second order in the curvature.
  General algorithms}}.
\bjtitle{Nucl. Phys. B}
\bvolume{333},
\bfpage{471}--\blpage{511}
(\byear{1990}).
\doiurl{10.1016/0550-3213(90)90047-H}
\end{barticle}
\endbibitem

\bibitem{Hamber2013}
\begin{barticle}
\bauthor{\bsnm{Hamber}, \binits{H.W.}},
\bauthor{\bsnm{Toriumi}, \binits{R.}}:
\batitle{{Inconsistencies from a Running Cosmological Constant}}.
\bjtitle{Int. J. Mod. Phys. D}
\bvolume{22}(\bissue{13}),
\bfpage{1330023}
(\byear{2013})
{\href{https://arxiv.org/abs/1301.6259}{{arXiv:1301.6259}}}
{[hep-th]}.
\doiurl{10.1142/S0218271813300231}
\end{barticle}
\endbibitem

\bibitem{Dyer:2008hb}
\begin{barticle}
\bauthor{\bsnm{Dyer}, \binits{E.}},
\bauthor{\bsnm{Hinterbichler}, \binits{K.}}:
\batitle{{Boundary Terms, Variational Principles and Higher Derivative Modified
  Gravity}}.
\bjtitle{Phys. Rev. D}
\bvolume{79},
\bfpage{024028}
(\byear{2009})
{\href{https://arxiv.org/abs/0809.4033}{{arXiv:0809.4033}}}
{[gr-qc]}.
\doiurl{10.1103/PhysRevD.79.024028}
\end{barticle}
\endbibitem

\bibitem{Corichi2022}
\begin{barticle}
\bauthor{\bsnm{Corichi}, \binits{A.}},
\bauthor{\bsnm{Nu\~{n}ez}, \binits{D.}}:
\batitle{{Introduction to the ADM formalism}}.
\bjtitle{Rev. Mex. Fis.}
\bvolume{37}(\bissue{4}),
\bfpage{720}--\blpage{747}
(\byear{1991})
{\href{https://arxiv.org/abs/arXiv:2210.10103v1
  [gr-qc]}{{arXiv:arXiv:2210.10103v1 [gr-qc]}}}.
\doiurl{10.48550/arXiv.2210.10103}
\end{barticle}
\endbibitem

\bibitem{Fukuma2001}
\begin{barticle}
\bauthor{\bsnm{Fukuma}, \binits{M.}},
\bauthor{\bsnm{Matsuura}, \binits{S.}},
\bauthor{\bsnm{Sakai}, \binits{T.}}:
\batitle{{Higher derivative gravity and the AdS / CFT correspondence}}.
\bjtitle{Prog. Theor. Phys.}
\bvolume{105},
\bfpage{1017}--\blpage{1044}
(\byear{2001})
{\href{https://arxiv.org/abs/hep-th/0103187}{{arXiv:hep-th/0103187}}}.
\doiurl{10.1143/PTP.105.1017}
\end{barticle}
\endbibitem

\bibitem{Gibbons1993}
\begin{bbook}
\bauthor{\bsnm{{Gibbons}}, \binits{G.W.}},
\bauthor{\bsnm{{Hawking}}, \binits{S.W.}}:
\bbtitle{{Euclidean Quantum Gravity}},
(\byear{1993})
\end{bbook}
\endbibitem

\bibitem{Delgado2022}
\begin{barticle}
\bauthor{\bsnm{Delgado}, \binits{R.C.}}:
\batitle{{Quantum gravitational corrections to the entropy of a
  Reissner\textendash{}Nordstr\"om black hole}}.
\bjtitle{Eur. Phys. J. C}
\bvolume{82}(\bissue{3}),
\bfpage{272}
(\byear{2022})
{\href{https://arxiv.org/abs/2201.08293}{{arXiv:2201.08293}}}
{[hep-th]}.
\doiurl{10.1140/epjc/s10052-022-10232-0}
\end{barticle}
\endbibitem

\bibitem{Xiao2022}
\begin{barticle}
\bauthor{\bsnm{Xiao}, \binits{Y.}},
\bauthor{\bsnm{Tian}, \binits{Y.}}:
\batitle{{Logarithmic correction to black hole entropy from the nonlocality of
  quantum gravity}}.
\bjtitle{Phys. Rev. D}
\bvolume{105}(\bissue{4}),
\bfpage{044013}
(\byear{2022})
{\href{https://arxiv.org/abs/2104.14902}{{arXiv:2104.14902}}}
{[gr-qc]}.
\doiurl{10.1103/PhysRevD.105.044013}
\end{barticle}
\endbibitem

\bibitem{Tsamis1995}
\begin{barticle}
\bauthor{\bsnm{Tsamis}, \binits{N.C.}},
\bauthor{\bsnm{Woodard}, \binits{R.P.}}:
\batitle{{Strong infrared effects in quantum gravity}}.
\bjtitle{Annals Phys.}
\bvolume{238},
\bfpage{1}--\blpage{82}
(\byear{1995}).
\doiurl{10.1006/aphy.1995.1015}
\end{barticle}
\endbibitem

\bibitem{Cardoso:2018ptl}
\begin{barticle}
\bauthor{\bsnm{Cardoso}, \binits{V.}},
\bauthor{\bsnm{Kimura}, \binits{M.}},
\bauthor{\bsnm{Maselli}, \binits{A.}},
\bauthor{\bsnm{Senatore}, \binits{L.}}:
\batitle{{Black Holes in an Effective Field Theory Extension of General
  Relativity}}.
\bjtitle{Phys. Rev. Lett.}
\bvolume{121}(\bissue{25}),
\bfpage{251105}
(\byear{2018})
{\href{https://arxiv.org/abs/1808.08962}{{arXiv:1808.08962}}}
{[gr-qc]}.
\doiurl{10.1103/PhysRevLett.121.251105}
\end{barticle}
\endbibitem

\bibitem{Cardoso:2019mqo}
\begin{barticle}
\bauthor{\bsnm{Cardoso}, \binits{V.}},
\bauthor{\bsnm{Kimura}, \binits{M.}},
\bauthor{\bsnm{Maselli}, \binits{A.}},
\bauthor{\bsnm{Berti}, \binits{E.}},
\bauthor{\bsnm{Macedo}, \binits{C.F.B.}},
\bauthor{\bsnm{McManus}, \binits{R.}}:
\batitle{{Parametrized black hole quasinormal ringdown: Decoupled equations for
  nonrotating black holes}}.
\bjtitle{Phys. Rev. D}
\bvolume{99}(\bissue{10}),
\bfpage{104077}
(\byear{2019})
{\href{https://arxiv.org/abs/1901.01265}{{arXiv:1901.01265}}}
{[gr-qc]}.
\doiurl{10.1103/PhysRevD.99.104077}
\end{barticle}
\endbibitem

\end{thebibliography}
\end{document}